\documentclass[sigconf, nonacm]{acmart}

\newcommand\vldbdoi{XX.XX/XXX.XX}
\newcommand\vldbpages{XXX-XXX}
\newcommand\vldbvolume{14}
\newcommand\vldbissue{1}
\newcommand\vldbyear{2020}
\newcommand\vldbauthors{\authors}
\newcommand\vldbtitle{\shorttitle} 
\newcommand\vldbavailabilityurl{URL_TO_YOUR_ARTIFACTS}
\newcommand\vldbpagestyle{plain} 

\usepackage{graphicx, subcaption, caption, wrapfig}
\usepackage{tcolorbox}
\usepackage{hyperref}       
\usepackage{url}            
\usepackage{booktabs}       
\usepackage{nicefrac}       
\usepackage{microtype}      
\usepackage{xcolor}         
\usepackage{enumitem}  
\usepackage{array}     
\usepackage{graphicx} 
\usepackage{makecell}
\usepackage{multirow}
\usepackage{colortbl}
\usepackage{pifont}
\usepackage{subcaption}  
\usepackage{caption}
\usepackage{pifont}
\usepackage{changepage}
\usepackage{float}
\usepackage{stfloats}
\usepackage{array}
\usepackage{tabularx}
\usepackage{wrapfig}
\usepackage{algorithm}
\usepackage{algorithmic}
\usepackage{fvextra}
\usepackage{multicol}
\usepackage{ragged2e}
\usepackage{titlesec}
\usepackage{placeins}

\tcbuselibrary{listings, breakable}

\definecolor{lightgreen}{RGB}{231, 243, 235}
\definecolor{darkgreen}{RGB}{104, 167, 82}

\definecolor{lightgray}{gray}{0.96}
\definecolor{codegray}{rgb}{0.3,0.3,0.3}

\newcommand{\paradigm}{\textit{Text-to-Pipeline} }
\newcommand{\sparadigm}{\textit{Text-to-Pipeline}}
\newcommand{\model}{\textsc{PARROT} }
\newcommand{\smodel}{\textsc{PARROT}}

\newcommand*{\images}[1]{\includegraphics[width=0.25cm,height=!]{#1}}

\newcommand{\vpara}[1]{\noindent \textbf{#1.}}

\DefineVerbatimEnvironment{myverbatim}{Verbatim}{breaklines=true}
\newtcolorbox{casebox}[1][]{colback=red!10, colframe=red!80!black,
  title=Case Study,#1}

\lstdefinestyle{code}{
  language=Python,
  basicstyle=\ttfamily\small,
  backgroundcolor=\color{gray!5},
  keywordstyle=\color{blue!70!black}\bfseries,
  stringstyle=\color{red!60!black},
  commentstyle=\color{gray!70},
  numberstyle=\tiny\color{gray!60},
  numbers=none,              
  showstringspaces=false,
  breaklines=true,
  frame=single,              
  rulecolor=\color{gray!40},
  frameround=tttt,           
  tabsize=2,
  belowskip=0.6em,
  aboveskip=0.6em,
  captionpos=b,
  keepspaces=true,
}

\begin{document}
\title{Text-to-Pipeline: Bridging Natural Language and Data Preparation Pipelines [Experiment, Analysis \& Benchmark]}
\author{Yuhang Ge}
\affiliation{%
  \institution{Zhejiang University}
}
\email{yuhangge@zju.edu.cn}

\author{Yachuan Liu}
\affiliation{%
  \institution{Zhejiang University}
}
\email{liuyachuan@zju.edu.cn}

\author{Zhangyan Ye}
\affiliation{%
  \institution{Zhejiang University}
}
\email{yzyet@zju.edu.cn}

\author{Yuren Mao}
\affiliation{%
  \institution{Zhejiang University}
}
\email{yuren.mao@zju.edu.cn}

\author{Yunjun Gao}
\affiliation{%
  \institution{Zhejiang University}
}
\email{gaoyj@zju.edu.cn}

\begin{abstract}
\noindent Data preparation (DP) transforms raw data into a form suitable for downstream applications, typically by composing operations into executable pipelines. Building such pipelines is time-consuming and requires sophisticated programming skills, posing a significant barrier for non-experts. To lower this barrier, we introduce \sparadigm, a new task that translates NL data preparation instructions into DP pipelines, and \smodel, a large-scale benchmark to support systematic evaluation. To ensure realistic DP scenarios, \model is built by mining transformation patterns from production pipelines and instantiating them on 23,009 real-world tables, resulting in \textasciitilde18,000 tasks spanning 16 core operators. Our empirical evaluation on \model reveals a critical failure mode in cutting-edge LLMs: they struggle not only with multi-step compositional logic but also with semantic parameter grounding. We thus establish a strong baseline with \textit{Pipeline-Agent}, an execution-aware agent that iteratively reflects on intermediate states. While it achieves state-of-the-art performance, a significant gap remains, underscoring the deep, unsolved challenges for \smodel. It provides the essential, large-scale testbed for developing and evaluating the next generation of autonomous data preparation agentic systems.
\end{abstract}

\maketitle

\pagestyle{\vldbpagestyle}
\begingroup\small\noindent\raggedright\textbf{PVLDB Reference Format:}\\
\vldbauthors. \vldbtitle. PVLDB, \vldbvolume(\vldbissue): \vldbpages, \vldbyear.\\
\href{https://doi.org/\vldbdoi}{doi:\vldbdoi}
\endgroup
\begingroup
\renewcommand\thefootnote{}\footnote{\noindent
This work is licensed under the Creative Commons BY-NC-ND 4.0 International License. Visit \url{https://creativecommons.org/licenses/by-nc-nd/4.0/} to view a copy of this license. For any use beyond those covered by this license, obtain permission by emailing \href{mailto:info@vldb.org}{info@vldb.org}. Copyright is held by the owner/author(s). Publication rights licensed to the VLDB Endowment. \\
\raggedright Proceedings of the VLDB Endowment, Vol. \vldbvolume, No. \vldbissue\ %
ISSN 2150-8097. \\
\href{https://doi.org/\vldbdoi}{doi:\vldbdoi} \\
}\addtocounter{footnote}{-1}\endgroup

\ifdefempty{\vldbavailabilityurl}{}{
\vspace{.3cm}
\begingroup\small\noindent\raggedright\textbf{PVLDB Artifact Availability:}\\
The source code, data, and/or other artifacts have been made available at \href{https://github.com/A11en0/Text-to-Pipeline}{https://github.com/A11en0/Text-to-Pipeline}.
\endgroup
}

\section{Introduction}
\label{sec:intro}

Data preparation~(DP) refers to the process of transforming raw data into a form suitable for downstream applications such as business intelligence~(BI) and machine learning~(ML)~\cite{chai2022data,fernandez2023large,zha2025data,zhang2023jellyfish}. As a core component of modern data management, tabular DP plays a central role in supporting workflows in data warehouses and BI systems~\cite{chai2022data}. Preparing tabular data typically involves multiple operations such as filtering~\cite{naeem2024retclean}, joining~\cite{deng2024lakebench}, grouping~\cite{auto-suggest}, and reshaping~\cite{he2018transform}. These operations are often composed into pipelines where each step incrementally transforms the table and feeds the result into the next~\cite{auto-pipeline,lai2025autoprep,auto-suggest}. However, building correct and efficient pipelines is time-consuming and requires sophisticated programming skills, which is challenging even for experienced data engineers, as they need to compose pipelines in a large compositional space. Moreover, this poses a significant technical barrier for non-experts and prevents them from participating in DP.
\begin{figure}
  \centering
  \includegraphics[width=1.0\linewidth]{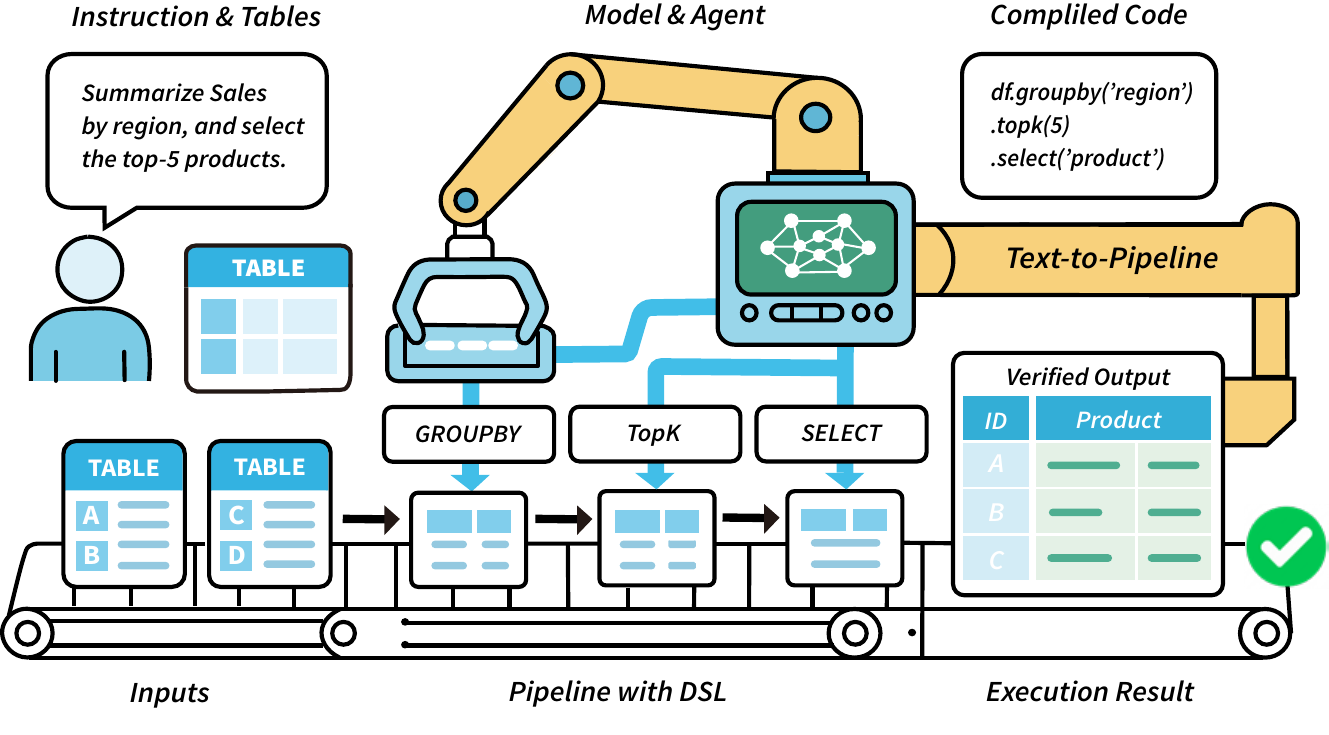}
  \caption{Task overview of \sparadigm.}
  \Description{}  
  \label{fig:workflow}
\end{figure}

To lower this barrier, a natural language~(NL) interface that allows users to complete DP tasks by writing NL instructions is a natural choice. This paradigm has been explored in related tasks such as Text-to-SQL~\cite{yu2018spider,lei2024spider}, spreadsheet formula generation~\cite{maspreadsheetbench,zhao2024nl2formula}, or code generation~\cite{wei2024selfcodealign,zhang2024codeagent}. However, these tasks primarily focus on database queries or cell-wise logic, not the multi-step, stateful transformations central to DP pipelines. In parallel, other automated pipeline construction methods~\cite{auto-pipeline,lai2025autoprep,auto-suggest} are not driven by NL; instead, they rely on structured supervision such as input-output table pairs~\cite{he2018transform,autopandas,auto-pipeline} or schema graphs~\cite{lai2025autoprep}, making them inapplicable when only NL instructions are available. This leaves a critical gap for NL-driven pipeline generation. Simply tasking LLMs to generate monolithic scripts in general-purpose languages like Pandas is also insufficient, as this approach lacks the modularity, formal verification, and robust state-tracking necessary to guarantee correctness over complex, multi-step transformations.

To bridge this gap, we introduce \paradigm task: translating NL instructions into executable DP pipelines over tabular data. We formalize it as symbolic program generation in a domain-specific language (DSL), which can be compiled into executable backend code such as Pandas or SQL. DSL offers a structured representation, backend flexibility, and stronger support for verification and evaluation than directly generating code. As shown in Fig.~\ref{fig:workflow}, an instruction ``\textit{Summarize sales by region, and select the top-5 products.}'' and input tables, the system generates a pipeline in DSL, e.g., {\texttt{GroupBy}, \texttt{Topk}, \texttt{Select}}, which is compiled into backend code and executed to produce the final table. To support this task, we build \smodel, a large-scale benchmark with \textasciitilde18,000 multi-step DP tasks spanning 16 core operators, using 23,009 real tables from six public sources. Constructing such a benchmark poses significant challenges: First, pipelines must be structurally realistic, reflecting real-world transformation patterns. Second, semantically valid (i.e., executable by construction on the given tables' schema). Third, faithfully aligned with diverse, natural NL instructions.

To address these challenges, we design a rigorous five-stage synthesis framework. We first extract transformation patterns from production pipelines to ensure our tasks align with practical usage. Next, we introduce a novel propose-then-validate synthesis process that separates structural realism from semantic validity. A \textit{Proposer}, guided by empirically-derived Markov transition matrices, samples structurally realistic operator chains. Then, a \textit{Validator}, powered by a formal Schema Propagation Mechanism (SPM), ensures each proposed operation is semantically valid given the current table state. This guarantees all generated pipelines are executable by construction. Finally, after compiling the DSL to executable code and generating intent-aligned NL using LLMs, we perform multi-phase validation, including a rigorous review by six PhD-level human experts, to ensure gold-standard data quality.

Our evaluation of \model validates our core hypothesis: formalism is critical. Targeting our symbolic DSL achieves \textbf{62.88\%} accuracy, vastly outperforming direct code generation like Pandas (\textbf{33.8\%}). Yet, even with a robust DSL, \textbf{even} cutting-edge LLMs like GPT-4o exhibit a critical failure: they achieve high program validity (\textbf{81.12\%}) but their execution accuracy drops to \textbf{71\%}, producing programs that \textit{run} but yield the \textit{wrong answer}. These findings highlight that direct prompting generation fails at the core challenges of multi-step compositional logic and semantic parameter grounding. To address this, we propose \textit{Pipeline-Agent}, an execution-aware agent that iteratively reasons over intermediate states, achieving a state-of-the-art performance of \textbf{76.17\%}. Despite this, the significant remaining gap underscores that these two challenges are the deep, unsolved problems posed by this task, calling for new solutions. In summary, our key contributions are:
\begin{itemize}[leftmargin=*]
    \item \textbf{Task:} We formalize the \paradigm task, translating natural language into executable, multi-step data preparation pipelines.
    \item \textbf{Benchmark:} We construct \smodel, a large-scale, high-quality benchmark of \textasciitilde18,000 instances, using a novel synthesis framework that guarantees pipeline validity and realism.
    \item \textbf{Findings:} We demonstrate that our DSL formalism is critical ($+29$ points over Pandas) and that SOTA LLMs fail on multi-step compositional logic and semantic parameter grounding.
    \item \textbf{Baseline:} We propose Pipeline-Agent, an SOTA agent that serves as a foundation for future research.
\end{itemize}
Finally, \model provides a challenging testbed for developing the next generation of intelligent data preparation agentic systems.



\section{Related Work}
\label{sec:related-work}

\vpara{NL-Driven Program Generation}
Prior research has extensively studied how to translate NL into executable programs. Specifically, Text-to-SQL methods~\cite{yu2018spider,lei2024spider,li2023bird,li2024codes} map NL queries to SQL statements, emphasizing semantic understanding and schema linking. These approaches are typically benchmarked on datasets such as WikiSQL~\cite{zhong2017seq2sql}, Spider~\cite{yu2018spider}, Spider 2.0~\cite{lei2024spider}, and BIRD~\cite{li2023bird}. Text-to-Formula~\cite{zhao2024nl2formula,li2023sheetcopilot,maspreadsheetbench} techniques like NL2Formula~\cite{zhao2024nl2formula}, SheetCopilot~\cite{li2023sheetcopilot}, and SpreadsheetBench~\cite{maspreadsheetbench} focus on converting NL into spreadsheet formulas or targeted cell-level edits, suitable for localized spreadsheet manipulations. Additionally, general Text-to-Code frameworks~\cite{feng2020codebert,chen2021evaluating,zhang2023survey,hendrycks2021measuring} translate NL instructions into scripts in languages like Python, C++, and Java, addressing diverse standalone programming tasks tested on HumanEval~\cite{chen2021evaluating} and APPS~\cite{hendrycks2021measuring}. Although these paradigms leverage NL, these tasks typically target the level of SQL queries, cell-wise formulas for spreadsheets, or logic functions for general programming. In contrast, \paradigm focuses on generating multi-step, executable pipelines for data preparation, where the objective is to transform input tables into expected outputs through schema-aware DSL programs.

\vpara{Automatic Data Pipeline Generation}
Automating data pipeline construction is often framed as program synthesis. Early methods rely on manual coding or visual tools~\cite{trifacta,power-query}, while example-driven approaches~\cite{he2018transform,gulwani2012spreadsheet,jin2017foofah,singh2016blinkfill,jin2018clx,heer2015predictive,barowy2015flashrelate,auto-join} require input-output~(IO) table pairs and struggle with multi-step logic. Subsequent work~\cite{autopandas,sql-by-example,auto-pipeline,query-by-output} extends to multi-step synthesis but still depends on output supervision. Auto-Tables~\cite{auto-tables} and Auto-Prep~\cite{lai2025autoprep} remove this constraint using self-supervised learning and mining existing operator traces, respectively. However, none of these methods support open-ended pipeline generation directly from NL instructions. In parallel, several human-interactive systems assist users during pipeline construction. EDAssistant~\cite{li2023edassistant} supports in-situ code search, and Auto-Suggest~\cite{auto-suggest} mines notebook patterns to recommend transformations. ChatPipe~\cite{chen2024chatpipe} enables conversational construction with LLMs. These systems rely on user interaction or code context, focusing on usability over automation. AutoPrep~\cite{fan2024autoprep} uses a multi-agent framework for question-aware preparation in TableQA, but does not support general-purpose pipeline generation. In contrast, our \paradigm task with \model targets fully automatic pipeline generation from NL instructions, enabling end-to-end generation without human feedback or IO supervision.

\section{Task Definition}
\label{sec:task-definition}

We formally define the \paradigm task as translating a natural language instruction over an input table set into an executable, multi-step data preparation pipeline. We represent this pipeline not as direct code, but as a symbolic program in a domain-specific language (DSL). This DSL representation is backend-agnostic, allowing it to be compiled into various executable codes.

\vpara{Problem Setup}
Let $\mathcal{X}$ denote the space of all possible input table sets, $\mathcal{Y}$ the space of output tables, $\mathcal{L}$ the space of natural language instructions, and $\mathcal{P}$ the space of DSL programs. An input instance consists of a set of one or more tables $\mathbf{x} = \{x_1, \dots, x_m\} \in \mathcal{X}$. Each instance is associated with a reference output table $y \in \mathcal{Y}$.
Given an instruction $\ell \in \mathcal{L}$ and an input table set $\mathbf{x} \in \mathcal{X}$, the objective is to learn a mapping $f: \mathcal{L} \times \mathcal{X} \rightarrow \mathcal{P}$ that yields a symbolic program $p = f(\ell, \mathbf{x})$, which can be compiled and executed to produce an output $\hat{y}$:
\begin{equation}
    \hat{y} = \texttt{Exec}(c, \mathbf{x}) \quad \text{where } c = \texttt{Compile}(p).
\end{equation}
Here, $\texttt{Compile}(p)$ is the compiled executable code, $\texttt{Exec}$ denotes the execution engine (e.g., Pandas or SQL), and $\hat{y}$ is expected to be equivalent to the reference output $y$. In summary, each instance of this \paradigm can be represented as a five-tuple $(\mathbf{x}, \ell, p, c, y)$.

\vpara{Task Scope}
The scope of \paradigm focuses on data preparation (e.g., cleaning, integration, transformation) over a given set of tables. This mirrors common enterprise workflows where data discovery is handled by upstream systems (e.g., data catalogs), and users then perform multi-step transformations on the discovered tables, as seen in tools like Trifacta~\cite{trifacta} or Power Query~\cite{power-query}.

\vpara{Pipeline Structure}
Each program \(p \in \mathcal{P}\) is  a left-to-right chain of \(k\) symbolic operators: $p = o_1 \circ o_2 \circ \dots \circ o_k$, where each operator $o_i \in \mathcal{O}$ is drawn from a core set covering common DP actions:
\begin{equation}
    \mathcal{O} = \{\texttt{groupby}, \texttt{sort}, \texttt{join}, \texttt{dropna}, \texttt{filter}, \dots\}.
\end{equation}
Each operator $o_i$ is parameterized by structured arguments (e.g., column names, aggregation functions). 
For a complete list of the 16 operators and their categories, refer to Tab.~\ref{tab:dsl-operators}. Despite its modular design, synthesizing such pipelines is non-trivial. The space of valid operator sequences grows combinatorially with length, and each step is constrained by schema compatibility, parameter validity, and cross-step dependencies. Moreover, subtle interactions among operators (e.g., column renaming before aggregation) can significantly affect program correctness and execution outcomes.

\vpara{Why DSL over Direct Code}
\label{para:dsl_vs_direct}
Compared to generating Pandas code directly, DSL offers three key advantages:
(1) \textit{Better planning and verification}: DSL supports step-wise reasoning, schema validation, and error tracing. It enables fine-grained evaluation (e.g., operator accuracy) and is easier to synthesize at scale. 
(2) \textit{Stable structure}: DSL uses a fixed set of operations with clear parameter formats, avoiding the syntactic variance of Pandas (e.g., multiple ways to filter or group data). This improves model learnability and consistency.
(3) \textit{Backend flexibility}: DSL can be compiled to Pandas, SQL, or Spark, making it adaptable to different runtime environments. In contrast, Pandas ties the output to Python execution. Experiments further confirm this choice, showing consistent performance advantages over direct Pandas and SQL generation (see Tab.~\ref{tab:target-generation-comparison}).

\vpara{Evaluation Metrics}
We employ three primary metrics to provide a multifaceted view of model performance, evaluating both execution correctness and program structure fidelity.
\begin{itemize}[leftmargin=*]
  \item \textbf{Execution Accuracy (EA)}: Given input tables $\mathbf{x}$ and a generated program $\hat{p}$, we execute $\hat{p}$ to obtain $\hat{y} = \texttt{Exec}(\hat{p}, x)$. EA measures the proportion of samples where $\hat{y} \overset{\star}{=} y$, i.e., the predicted output $\hat{y}$ matches the ground truth $y$ up to canonical equivalence (e.g., row/column permutations, floating-point tolerance):
  \[
  \text{EA} = \frac{1}{N} \sum_{i=1}^N \mathbb{I}(\hat{y}_i \overset{\star}{=} y_i),
  \]
  where $N$ is the number of test samples, and $\mathbb{I}(\cdot)$ is the indicator function.  
  \item \textbf{Program Validity (PV)}: The proportion of generated programs $\hat{p}$ that can be successfully compiled into executable code $c = \texttt{Compile}(\hat{p})$ \textit{and} be executed ($ \hat{y} = \texttt{Exec}(c, \mathbf{x}) $) without any runtime errors, regardless of output correctness:
  \[
  \text{PV} = \frac{1}{N} \sum_{i=1}^N \mathbb{I}(\texttt{Valid}(\hat{p}_i)),
  \]
  where $\texttt{Valid}(\hat{p}_i)$ returns true if $\hat{p}_i$ compiles and executes successfully.  
  \item \textbf{Operator Accuracy (OA)}: This metric provides a complementary, order-agnostic view, measuring the proportion of correctly identified operators. It evaluates the selection of the correct \textit{set} of operations, while EA evaluates the sequence. For a generated program $\hat{p}$ and ground truth $p$, OA is computed as:
  \[
  \text{OA} = \frac{1}{N} \sum_{i=1}^N \frac{|\text{set}(\hat{p}_i) \cap \text{set}(p_i)|}{|\text{set}(p_i)|},
  \]
  where $\text{set}(p_i)$ denotes the set of operators in the ground truth program $p_i$.
\end{itemize}

\section{Benchmark Design}
\label{sec:benchmark-design}
\begin{figure*}
    \centering
    \includegraphics[width=1.0\textwidth]{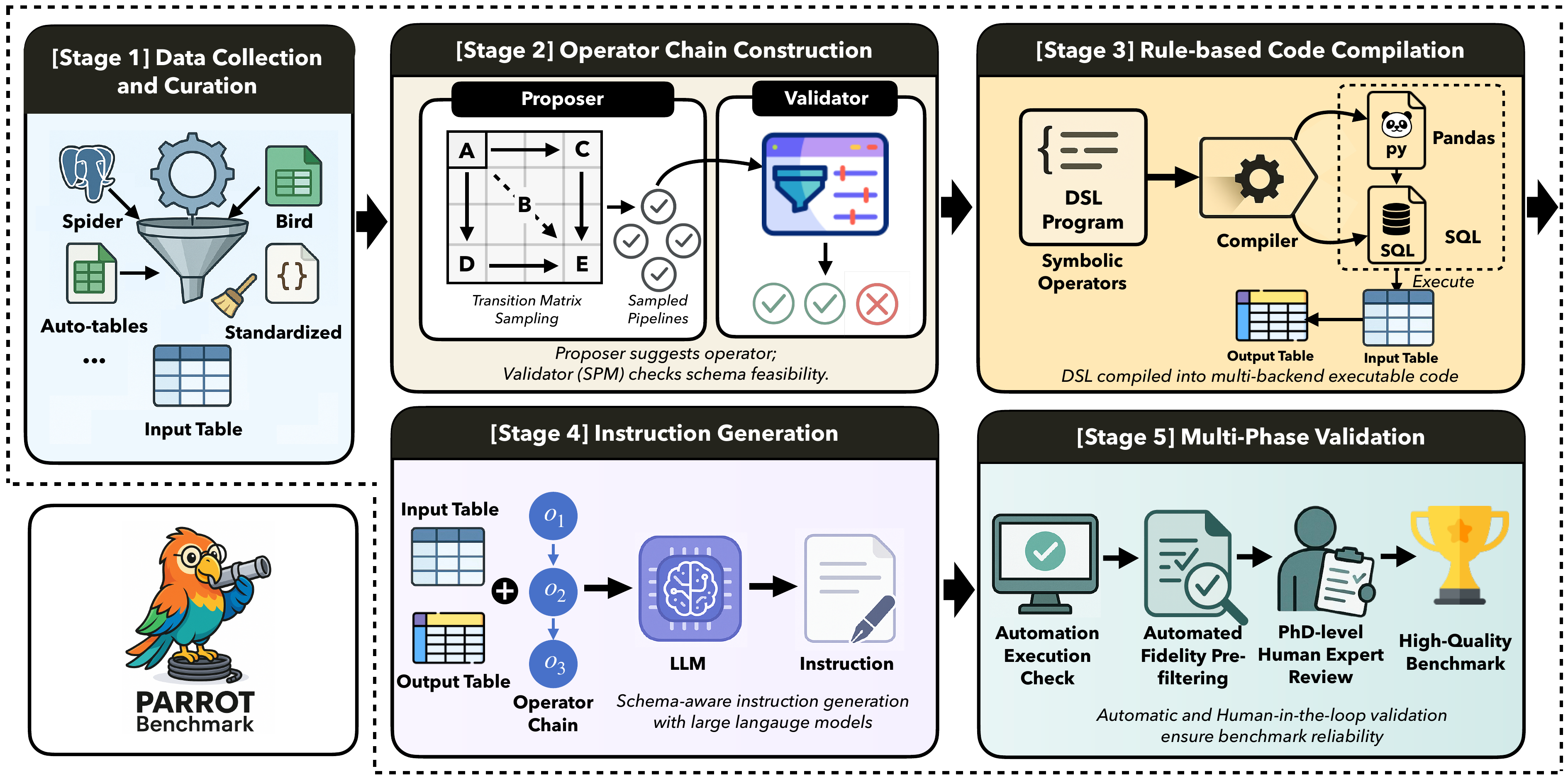}
    \caption{The data synthesis workflow of \smodel.}
    \label{fig:framework}
\end{figure*}

To support systematic evaluation of \sparadigm, we synthesize a high-quality benchmark of about 18,000 instances through a carefully designed data construction framework. As shown in Fig.~\ref{fig:framework}, the process involves five stages:
(1) \textbf{Data Collection and Curation}, where we gather and standardize tables from diverse real-world sources; 
(2) \textbf{Operator Chain Construction}, where we sample realistic, schema-valid transformation pipelines using a Markov model guided by our Schema Propagation Mechanism (SPM); 
(3) \textbf{Rule-Based Code Compilation}, where we deterministically compile the abstract pipelines into executable code to generate the ground-truth output tables; 
(4) \textbf{Instruction Generation}, where we use large language models to create natural language instructions corresponding to the pipelines and their I/O tables; 
and (5) \textbf{Multi-phase Validation}, where we employ both automatic execution checks and human-in-the-loop verification to ensure data quality. We detail each of these five stages in the following subsections.

\subsection{Stage 1: Data Collection and Curation}
\label{sec:data-collection-curation}
To simulate realistic DP scenarios, we collect tables from six public datasets covering diverse real-world domains.
\begin{itemize}[leftmargin=*]
    \item \textbf{Spider}~\cite{yu2018spider} is a large-scale, human-annotated benchmark for Text-to-SQL, designed to test cross-domain generalization. It features 200 databases with multiple tables, requiring models to handle complex queries and intricate schema relationships.
    \item \textbf{Bird}~\cite{li2023bird} is a large-scale benchmark for Text-to-SQL focused on real-world business intelligence (BI) scenarios. It includes large databases with varied, noisy table structures and complex join conditions, mirroring challenges in enterprise environments.
    \item \textbf{Auto-Pipeline}~\cite{auto-pipeline} and \textbf{Auto-Tables}~\cite{auto-tables} are benchmarks designed to advance the automation of complex data preparation workflows. They focus on synthesizing multi-step transformations, such as joins, pivots, and unpivoting, often without user-provided examples (e.g., using a ``by-target'' paradigm).
    \item \textbf{TableBench}~\cite{wu2025tablebench} is a comprehensive benchmark for Table Question Answering (TableQA) that emphasizing complex reasoning across diverse domains such as fact checking, numerical reasoning, and data analysis.
    \item \textbf{LakeBench}~\cite{deng2024lakebench} is a massive-scale (over 1TB) benchmark designed to evaluate methods for discovering joinable and unionable tables within data lakes. It features over 10,000 queries with ground truth annotations.
\end{itemize}

From these sources, we curated 23,009 tables, spanning domains such as blockchain, finance, healthcare, and education. These tables exhibit diverse structures, ranging from 2 to 120 columns, and include both wide and long formats. To ensure downstream compatibility, we then performed several curation steps to ensure quality and tractability. First, we performed structural validation, filtering out any empty or irregular tables that lacked a coherent schema. Second, to ensure manageable execution times for our benchmark tasks, we enforced a uniform row limit, truncating tables that exceeded 50 rows while preserving their original schema. Finally, we applied basic preprocessing to clean superficial data inconsistencies. This included steps such as trimming leading/trailing whitespace from string values and unifying common null representations (e.g., literal strings like ``NA'', ``Null'', or empty strings ``'') into a standard NaN marker. This curation process resulted in a robust and diverse collection of tables used for the pipeline synthesis stage.
\begin{table*}[t]
  \centering
  \caption{Supported operators: categorized by task type, typical parameters, Pandas-style examples, and observed frequencies.}
  \renewcommand{\arraystretch}{0.9}
  \small
  \resizebox{1.0\linewidth}{!}{
  \begin{tabular}{llll}
  \toprule
  \textbf{Operator} & \textbf{Typical Parameters} & \textbf{Example (Pandas-style)} & \textbf{Frequency} \\
  \midrule
  \multicolumn{4}{l}{\textbf{Data Cleaning}} \\
  \texttt{filter}         & condition             & \texttt{df.query("value != 1")} & 8876 \\
  \texttt{dropna}         & axis, how, subset     & \texttt{df.dropna()} & 3925 \\
  \texttt{deduplicate}    & subset, keep          & \texttt{df.drop\_duplicates()} & 7065 \\
  \texttt{cast}           & column, dtype         & \texttt{df["value"].astype("float")} & 3888 \\
  \midrule
  \multicolumn{4}{l}{\textbf{Data Integration}} \\
  \texttt{join}           & left, right, on, how  & \texttt{df1.merge(df2, on="id", how="inner")} & 3643 \\
  \texttt{union}          & dataframes            & \texttt{pd.concat([df1, df2])} & 2198 \\
  \midrule
  \multicolumn{4}{l}{\textbf{Structural Reconstruction}} \\
  \texttt{groupby}        & by, agg               & \texttt{df.groupby("region").sum()} & 9271 \\  
  \texttt{pivot}          & index, columns, values & \texttt{df.pivot("id", "type", "score")} & 2933 \\
  \texttt{unpivot}        & id\_vars, value\_vars  & \texttt{df.melt(id\_vars=["id"], value\_vars=["value"])} & 4365 \\
  \texttt{explode}        & column                & \texttt{df.explode("tags")} & 2135 \\
  \texttt{transpose}      & -                     & \texttt{df.transpose()} & 2271 \\
  \texttt{wide\_to\_long} & stubnames, i, j       & \texttt{pd.wide\_to\_long(...)} & 588 \\
  \midrule
  \multicolumn{4}{l}{\textbf{Assisted Operations}} \\
  \texttt{sort}           & by, ascending         & \texttt{df.sort\_values("time")} & 8588 \\
  \texttt{topk}           & columns, k            & \texttt{df.head(5)} & 4074 \\
  \texttt{select}         & columns               & \texttt{df[["name", "value"]]} & 5762 \\
  \texttt{rename}         & columns               & \texttt{df.rename(columns=\{"old": "new"\})} & 5937 \\
  \bottomrule
  \end{tabular}}
  \label{tab:dsl-operators}
\end{table*}

\subsection{Stage 2: Operator Chain Construction}
\label{sec:operator-chain-construction}

This stage aims for synthesizing operator chains $p = o_1 \circ \dots \circ o_k$ that are both structurally realistic and semantically valid. 
Our synthesis framework is designed as a propose-then-validate process.
At each step of building the chain, this process performs two key actions:
(1) \textbf{Propose}: the \textit{Proposer}, trained on real-world operator sequences, suggests a set of structurally plausible next operators.
(2) \textbf{Validate}: the \textit{Validator}, powered by a formal schema engine, checks if a proposed operator is semantically executable given the current table schema. This iterative propose-then-validate loop guarantees that every synthesized pipeline $p$ is executable by construction. We now detail the design of these two core components.

\vspace{0.5em}
\vpara{The Proposer}
The Proposer's role is to ensure that generated operator sequences reflect common patterns found in real-world data preparation. To achieve this, we model operator transitions as a first-order Markov process. We first define our DSL with 16 core operators (see Tab.~\ref{tab:dsl-operators}). We then construct an empirical transition matrix $\mathbf{P} \in \mathbb{R}^{|\mathcal{O}| \times |\mathcal{O}|}$. To build this matrix, we collected and parsed 1,200 data transformation scripts from open-source repositories (e.g., Kaggle, GitHub)~\cite{auto-tables,auto-pipeline,auto-suggest}. Each script was parsed into an abstract syntax tree (AST), and high-level transformation calls were extracted and mapped to our normalized DSL operator set (e.g., \texttt{df.drop\_duplicates} $\rightarrow$ \texttt{dedup}). We then recorded all observed consecutive operator pairs $(o_i \rightarrow o_j)$ to compute transition frequencies. These frequencies were normalized into conditional probabilities $P(o_j \mid o_i)$ using Laplace smoothing ($\alpha = 0.5$) to mitigate data sparsity. This resulting matrix $\mathbf{P}$ serves as our generative model for proposing structurally realistic operators.

\vspace{0.5em}
\vpara{The Validator}
The Proposer (using $\mathbf{P}$) only ensures structural realism; it has no knowledge of schema and may propose an invalid operation (e.g., applying \texttt{groupby} to a non-existent column). The Validator's role is to prevent this. 
We designed the Validator around a core engine we call the \textbf{Schema Propagation Mechanism (SPM)}. The SPM formally defines the effect of any operator $o_t$ on a given schema $\mathcal{S}_t$ at each step $t$, producing a new output schema $\mathcal{S}_{t+1}$. 
Specifically, the schema is represented as a structured object $\mathcal{S}_t = \{ (c_i, \tau_i) \}_{i=1}^{n}$, where $c_i$ is the column name and $\tau_i$ is its data type. 
Each DSL operator is associated with a deterministic transformation rule $\delta: \mathcal{S}_t \rightarrow \mathcal{S}_{t+1}$ that specifies how it modifies the schema. For example, \texttt{groupby} introduces new aggregation columns and may drop non-grouped fields; \texttt{join} merges two schemas, applying name disambiguation (e.g., suffixes \texttt{\_x}, \texttt{\_y}) if overlapping columns exist; \texttt{rename} updates column names while preserving types; \texttt{pivot} restructures column names based on index and values; and \texttt{dropna} does not modify the schema structure.
During chain construction, we maintain a running schema $\mathcal{S}_t$, updating it at each step via $\mathcal{S}_{t+1} = \delta(o_t, \mathcal{S}_t)$. This propagation strategy enables robust error prevention: before an operator $o_t$ is added to the chain, the Validator checks if its required arguments (e.g., column names) are present and well-typed in $\mathcal{S}_t$. If the check fails, the operator is rejected. For multi-source operations (e.g., \texttt{join}, \texttt{union}), the SPM tracks a set of schemas $\{\mathcal{S}_t^{(1)}, \mathcal{S}_t^{(2)}, \dots\}$ and applies compatibility checks, such as matching join keys.

\vspace{0.5em}
\vpara{Schema-Aware Synthesis Algorithm}
We now combine the Proposer and Validator into a unified algorithm, as summarized in Algo.~\ref{algo:chain-construction}. The initial operator $o_1$ is drawn uniformly. At each subsequent step $i$, the Proposer \textit{proposes} a candidate operator $o_i$ by sampling from $\mathbf{P}(\cdot \mid o_{i-1})$. This candidate is then passed to the Validator (SPM) for verification.

This validation step is critical: the SPM not only checks if $o_i$ is applicable but also guides the selection of valid parameters (e.g., columns) for $o_i$. If the operator is successfully validated and parameterized, it is appended to the chain $p$, and the SPM updates the schema via $\mathcal{S}_i = \delta(o_i, \mathcal{S}_{i-1})$. If validation fails (e.g., no valid columns can be found), the operator $o_i$ is simply discarded, and the algorithm proceeds to the next iteration $i+1$.
The chain length $k$ is sampled from a truncated geometric distribution over $[1, 8]$. We also adopt a three-level difficulty scheme based on program length (e.g., Easy ($\leq$ 3 ops), Medium (4-6 ops), and Hard ($\geq$7 ops)), as detailed in Sec.~\ref{sec:benchmark-characteristic}. This entire process guarantees that all synthesized pipelines $p$ are executable by construction. The detailed definitions and illustrative examples for each level are provided in App.~\ref{app:difficulty}.
\begin{algorithm}[t]
\caption{Operator Chain Construction}
\label{algo:chain-construction}
\begin{algorithmic}[1]
\REQUIRE Curated table set $\mathbf{x}$, Transition Matrix $\mathbf{P}$, Schema Propagation Mechanism (SPM)
\ENSURE Valid operator chain $p = \{o_1, o_2, \dots, o_k\}$

\STATE Sample chain length $k \sim \text{TruncGeom}(1, 8)$
\STATE Sample first operator $o_1$ uniformly from $\mathcal{O}$
\STATE Initialize chain $p \leftarrow [o_1]$
\STATE Get initial schema $\mathcal{S}_0 \leftarrow \texttt{GetSchema}(\mathbf{x})$
\FOR{$i = 2$ to $k$}
    \STATE // Propose a structurally realistic operator
    \STATE Sample $o_i \sim \mathbf{P}(\cdot \mid o_{i-1})$ 
    
    \STATE // Validate operator and bind parameters using SPM
    \STATE $is\_valid, o_i \leftarrow \texttt{SPM.ValidateAndBind}(o_i, \mathcal{S}_{i-1})$
    
    \IF{$is\_valid$ is \texttt{True}}
        \STATE Append $o_i$ to chain $p$
        \STATE $\mathcal{S}_i \leftarrow \texttt{SPM.UpdateSchema}(\mathcal{S}_{i-1}, o_i)$ // Propagate schema
    \ENDIF
    \STATE // If not valid, operator is discarded and loop continues
\ENDFOR
\RETURN $p$
\end{algorithmic}
\end{algorithm}

\subsection{Stage 3: Rule-Based Code Compilation}
\label{sec:code-compilation}
After yields a validated symbolic pipeline $p$, we compile it into executable code $c$. This serves two purposes: (1) generating the final output table $y$ (by executing $c$), which is required for instruction generation in Stage~\ref{sec:instruction-generation}, and (2) creating the executable ground truth for evaluation. Since $p$ has been validated by the SPM, the compiler can be a simple, deterministic, and stateless template engine.
 
\vpara{Compilation workflow}
We implement a compiler to translate DSL programs into executable code. The process maps each operator to backend-specific templates, ensuring correct parameter binding, schema handling, and code generation. Given a DSL program $p = [o_1, o_2, \dots, o_k]$, the compiler performs the following:
\begin{enumerate}[leftmargin=*]
    \item \textbf{Parameter binding:} For each operator $o_i$, extract its parameters and map them to template slots;
    \item \textbf{Schema-aware quoting:} Apply column name escaping (e.g., \texttt{\_quote(col)}) to handle special characters;
    \item \textbf{Code merging:} Concatenate code lines using a left-to-right chaining convention (e.g., method chaining in Pandas);
    \item \textbf{Execution trace annotation:} Optionally insert line comments to denote DSL operator source (useful for debugging).
\end{enumerate}

\vpara{Backend support}
While Pandas serves as the primary backend, the compiler is modular and supports alternate targets such as SQL or Spark through switchable template registries. Each backend maintains an operator-template mapping, enabling flexible deployment without modifying the symbolic layer.

\vpara{Example}
Given the DSL sequence:
\begin{lstlisting}[style=code, label={lst:groupby_sort_dsl}, caption={}]
[
  {
    "op": "groupby",
    "params": {
      "by": ["region"],
      "agg": { "sales": "sum" }
    }
  },
  {
    "op": "sort",
    "params": { "by": "sales", "ascending": false }
  }
]
\end{lstlisting}

The compiler generates the following Pandas code:
\begin{lstlisting}[style=code, label={lst:groupby_sort_pandas}, caption={}]
df.groupby('region')['sales']
    .sum()
    .reset_index()
    .sort_values(by='sales', ascending=False)
\end{lstlisting}
This translation preserves the semantics of the symbolic pipeline while ensuring correctness and interpretability.

\subsection{Stage 4: Instruction Generation}
\label{sec:instruction-generation}
With the symbolic pipeline $p$ from Stage 2 and the corresponding output table $y$ (generated by executing the compiled code from Stage 3), the next step is to generate a natural language instruction $\ell$. Using the operator chain and an input-output table preview, $\{\mathbf{x}, y, p\}$, we generate instructions via a two-step process designed to ensure both semantic fidelity and linguistic diversity.

First, we prompt an LLM to generate a schema-aware, structured pipeline description grounded in the transformation. This initial prompt includes the task description, table schema, sampled rows, and in-context demonstrations to anchor the operator-language mappings and ensure semantic preservation, that is, making sure the generated text accurately reflects the underlying operations. Second, this structured draft is passed to a style-controlled refinement step, which converts the draft into a fluent, user-centric instruction. This two-step process is crucial as it decouples the goal of semantic grounding (Step 1) from that of stylistic expression (Step 2), which helps avoid templating bias.

For this entire process, we use \texttt{GPT-4o-mini} via OpenAI's API (\texttt{gpt-4o-mini-2024-07-18}) with a temperature of 0.7. This non-zero temperature is intentionally set to encourage linguistic diversity and mitigate the risk of deterministic, model-specific phrasing. The effectiveness of this de-biasing strategy is empirically validated in our benchmark analysis (see Sec.~\ref{sec:benchmark-characteristic}, Tab.~\ref{tab:diversity-metrics}), where \model demonstrates significantly higher lexical diversity (Distinct-n) and lower redundancy (Self-BLEU) than prior benchmarks. All prompts used for data synthesis are listed in App.~\ref{app:data-synthesis-prompts}.

\subsection{Stage 5: Multi-phase Validation}
\label{sec:multi-phase-validation}
To ensure the quality and reliability of \smodel, every generated instance $(\mathbf{x}, \ell, p, c, y)$ undergoes a rigorous multi-phase validation process, combining automated checks with expert human review.

\vpara{Phase 1: Automatic Execution Validation}
First, we perform an automatic execution check. The compiled code $c$ is executed on the input tables $\mathbf{x}$ to produce a predicted output $\hat{y}$. This $\hat{y}$ is then compared against the ground-truth $y$ for canonical equivalence, denoted $\hat{y} \overset{\star}{=} y$.
This check is robust to superficial formatting and ordering differences. Specifically, it verifies equivalence by: (1) sorting both tables by all columns to make row order invariant, (2) re-indexing columns to make column order invariant, and (3) applying a floating-point tolerance (e.g., $10^{-5}$) for all numeric fields. Any instance where $\hat{y} \overset{\star}{=} y$ evaluates to \texttt{False} is discarded.
    
\vpara{Phase 2: Automated Fidelity Pre-filtering}
Second, we assess instruction fidelity. We employ an LLM-based "judge" to pre-filter samples, checking the semantic alignment between the natural language instruction $\ell$ (from Stage 4) and the symbolic operator chain $p$ (from Stage 2). Using zero-shot classification prompts, the judge LLM scores the semantic consistency. Samples with low computed alignment scores are automatically discarded, reducing the burden on human annotators.

\vpara{Phase 3: Expert Manual Validation (for Dev/Test Sets)}
Finally, to build a gold-standard evaluation set, we manual review of the data intended for the development and test sets, ensuring the highest possible data quality.

\begin{itemize}[leftmargin=*]
    \item \textbf{Sampling:} We employed a stratified sampling strategy to select 3,000 instances for this review, drawing samples proportionally across both the three difficulty levels (Easy, Medium, Hard) and the 16 core operators. This ensures the reviewed subset mirrors the full benchmark's distribution.   
    \item \textbf{Process:} We recruited and trained six graduate-level experts (MSc/PhD students in data science and NLP) with detailed annotation guidelines. Each expert scored instances on a 3-point scale across three criteria: (1) Instruction Accuracy (faithfully reflects the pipeline?), (2) Operator Coverage (all key steps implied?), and (3) Semantic Clarity (unambiguous?).
    \item \textbf{Outcome:} The inter-annotator agreement reached 91.4\% (Cohen’s $\kappa$ = 0.82), confirming high reliability. Only samples that passed all automated checks (Phases 1 \& 2) and received unanimous approval on all three criteria from our experts were retained for the final development and test sets. The remaining set of automatically validated data serves as the training set. To facilitate this rigorous human-in-the-loop process, we developed an interactive visualization platform to assist experts in inspecting and debugging data, which is detailed in App.~\ref{app:platform}.
\end{itemize}

\begin{table*}[tbp]
\centering
\caption{Benchmark comparison across different tasks and execution properties. \model supports multi-step DSL programs grounded in natural language, with execution verified across multiple backends. Task types: TS = Text-to-SQL, TF = Text-to-Formula, P = Pipeline Synthesis, TP = Text-to-Pipeline. The symbol `-' indicates that it does not provide codes or instructions.}
\label{tab:benchmark-comparison}
\resizebox{\linewidth}{!}{
\begin{tabular}{lccccccccc}
\toprule
\textbf{Benchmark} & \textbf{Task Type} & \textbf{\#Inst.} & \textbf{\#Tabs} & \textbf{\#Ops} & \textbf{Avg. Steps} & \textbf{Atomic~Exec}                  & \textbf{NL-Driven}                    & \textbf{Multi-Backend}  \\ 
\midrule
Spider~\cite{yu2018spider}            & TS            & 10,181           & 1,056            & 27             & 4.78                & \images{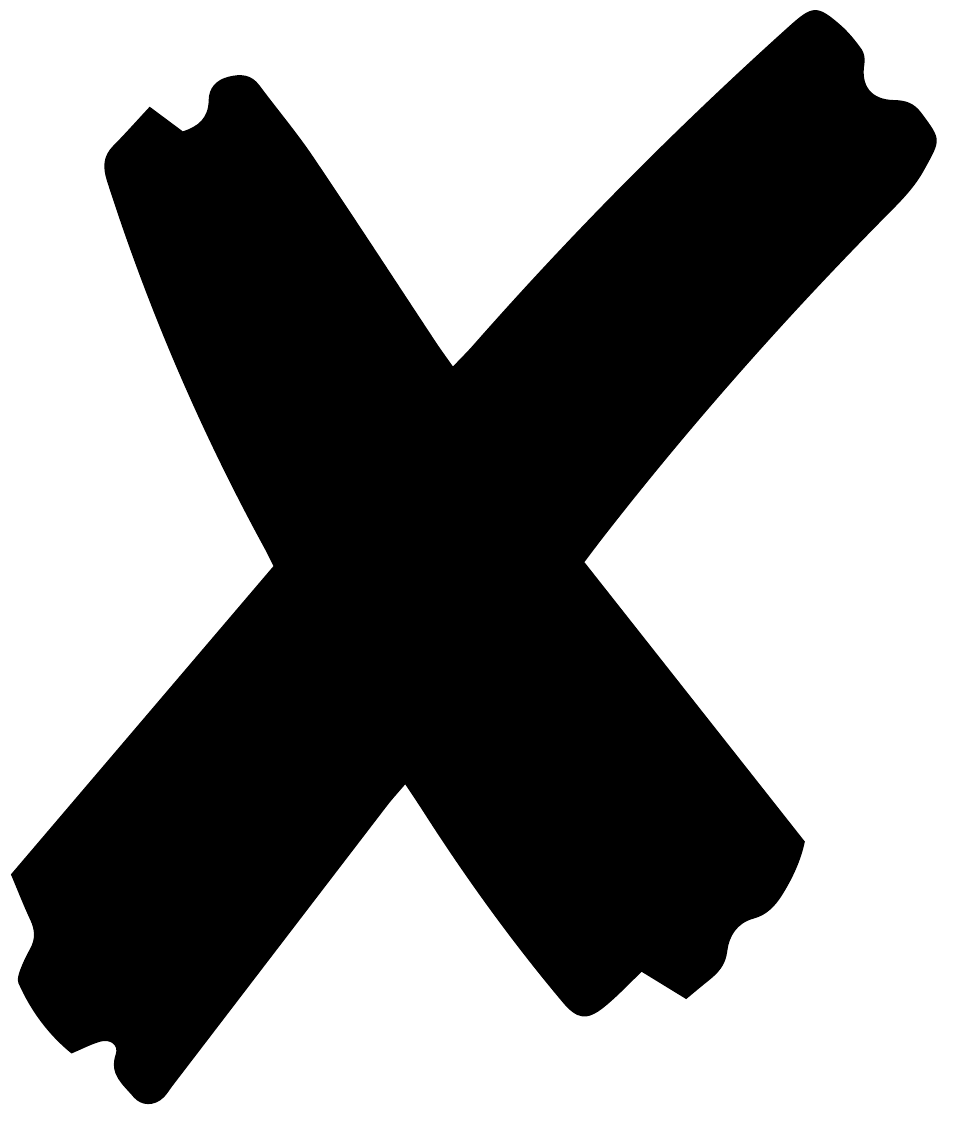} & \images{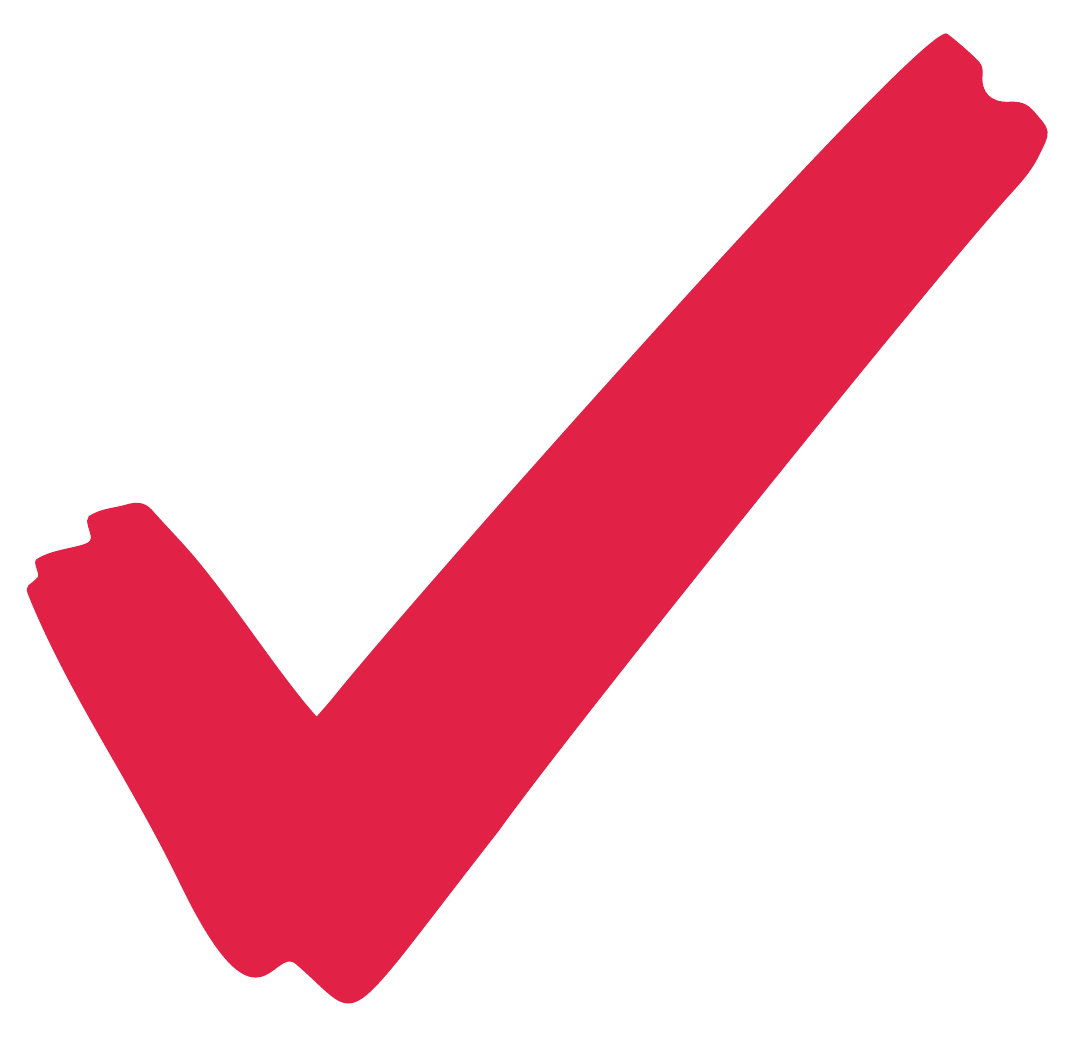} & \images{figures/not}             \\
BIRD~\cite{li2023bird}              & TS            & 12,751           & 693               & 27             & 6.52                & \images{figures/not} & \images{figures/yes} & \images{figures/not}             \\
SpreadsheetBench~\cite{maspreadsheetbench}  & TF            & 912              & 2,729             & 10             & -                   & \images{figures/not} & \images{figures/yes} & \images{figures/not}            \\
NL2Formula~\cite{zhao2024nl2formula}         & TF            & 70799              & 21670              & 57             & 10.2                   & \images{figures/not} & \images{figures/yes} & \images{figures/not}             \\
Auto-Tables~\cite{auto-tables}       & P             & 244              & 244               & 8              & 1.11                & \images{figures/yes} & \images{figures/not} & \images{figures/yes}             \\
Auto-Pipeline~\cite{auto-pipeline}     & P             & 716              & 4,680             & 12             & 4.10                   & \images{figures/yes} & \images{figures/not} & \images{figures/yes}             \\
\textbf{\smodel (Ours)}     & TP            & 17,168           & 23,009             & 16             & 4.24  & \images{figures/yes} & \images{figures/yes} & \images{figures/yes}             \\
\bottomrule
\end{tabular}}
\end{table*}

\section{Benchmark Characteristics}
\label{sec:benchmark-characteristic}

We present a benchmark comparison in Tab.~\ref{tab:benchmark-comparison} and a statistical overview in Tab.~\ref{tab:benchmark-stats}. Compared with prior works, \model provides large-scale, NL-driven multi-step programs with broad operator coverage, verified atomic execution, and multi-backend support.

\subsection{Compositional and Parameter Complexity}
The complexity of \model is two-fold: the \textit{compositional} depth of the programs and the \textit{parametric} complexity of individual operators.

\begin{figure*}[tbp]
  \centering
  \begin{minipage}[t]{0.32\linewidth}
      \centering
      \includegraphics[width=\linewidth]{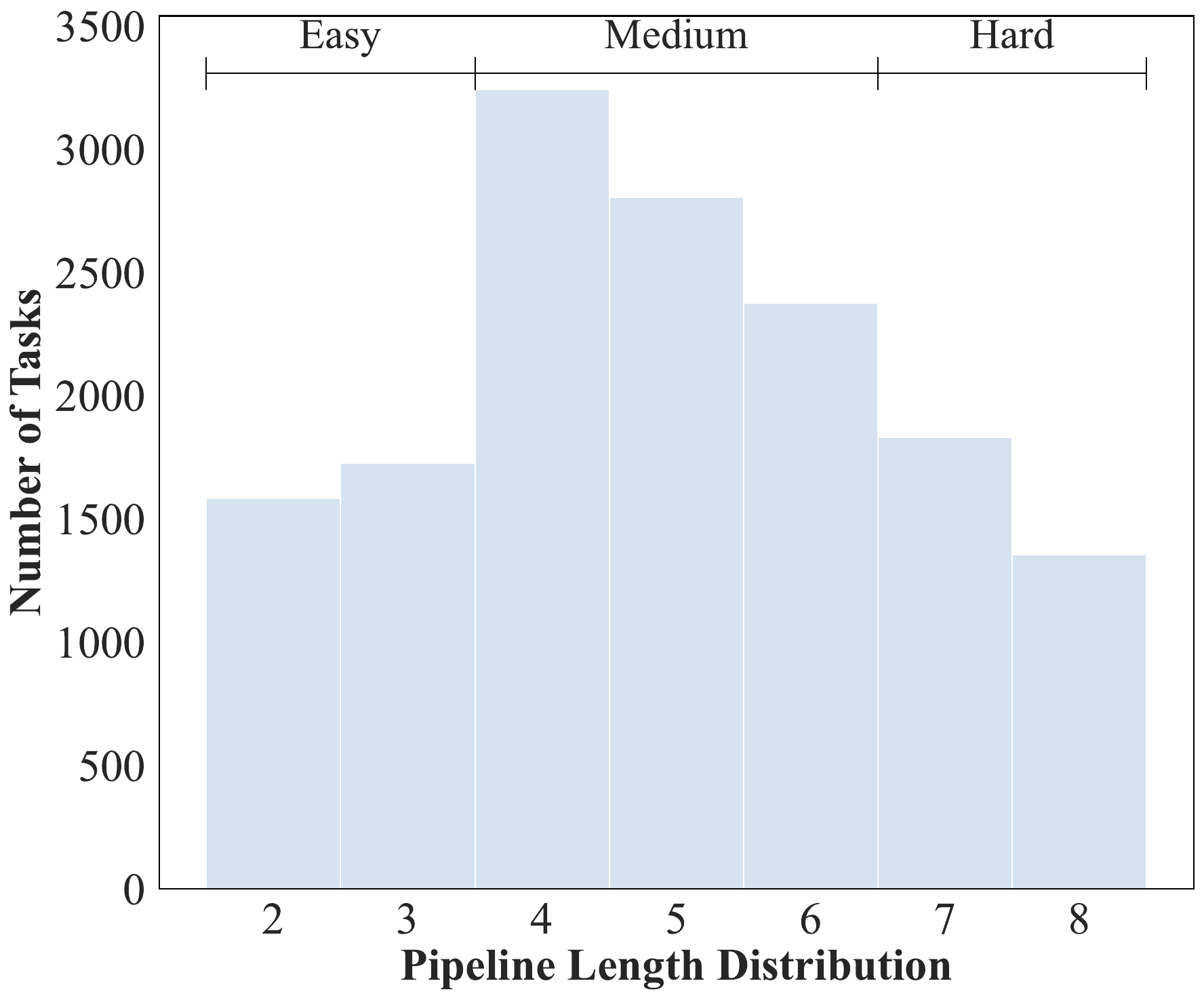}
  \end{minipage}
  \hfill 
  \begin{minipage}[t]{0.32\linewidth}
      \centering
      \includegraphics[width=\linewidth]{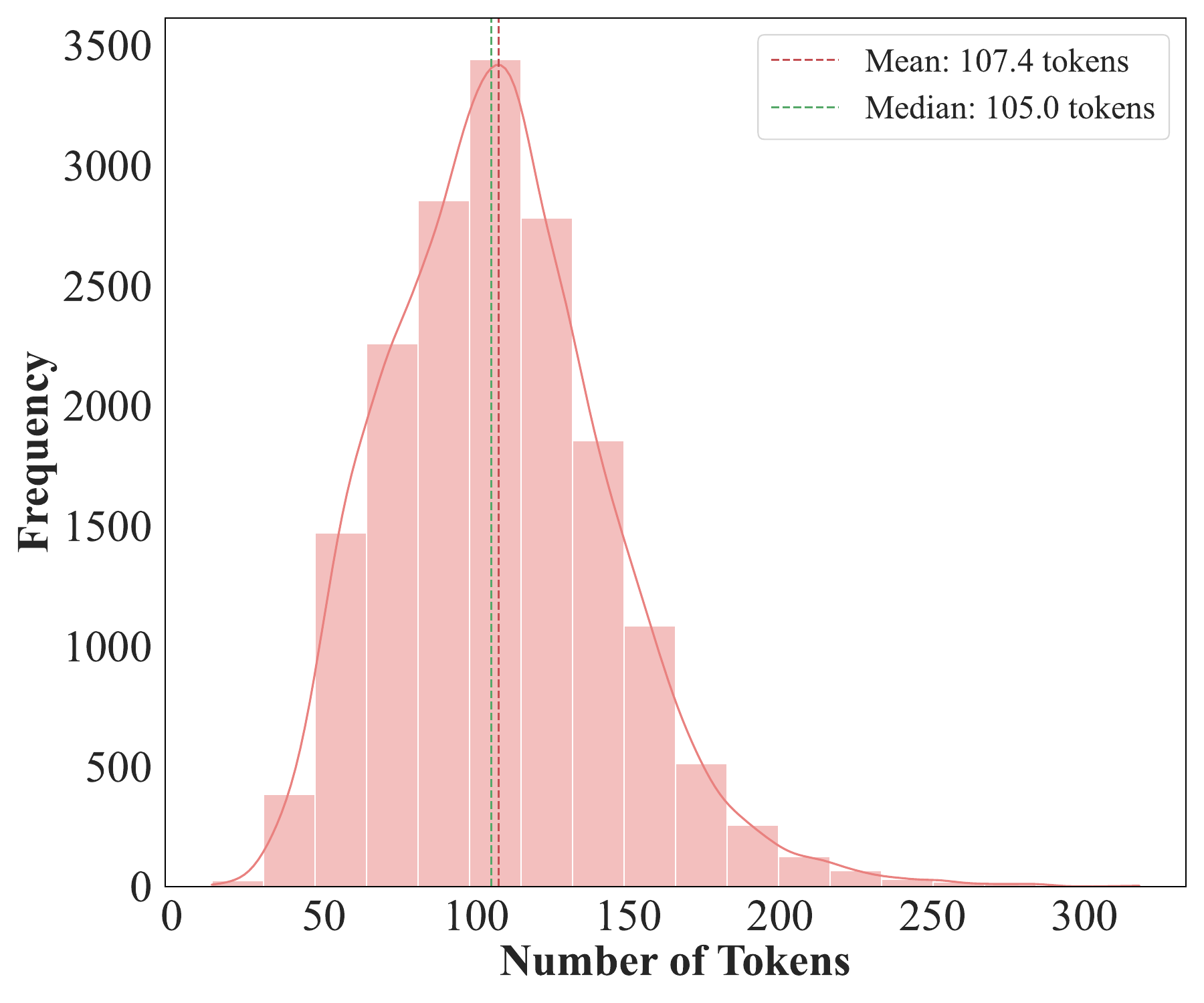}
  \end{minipage}
  \hfill
  \begin{minipage}[t]{0.33\linewidth}
      \centering
      \includegraphics[width=\linewidth]{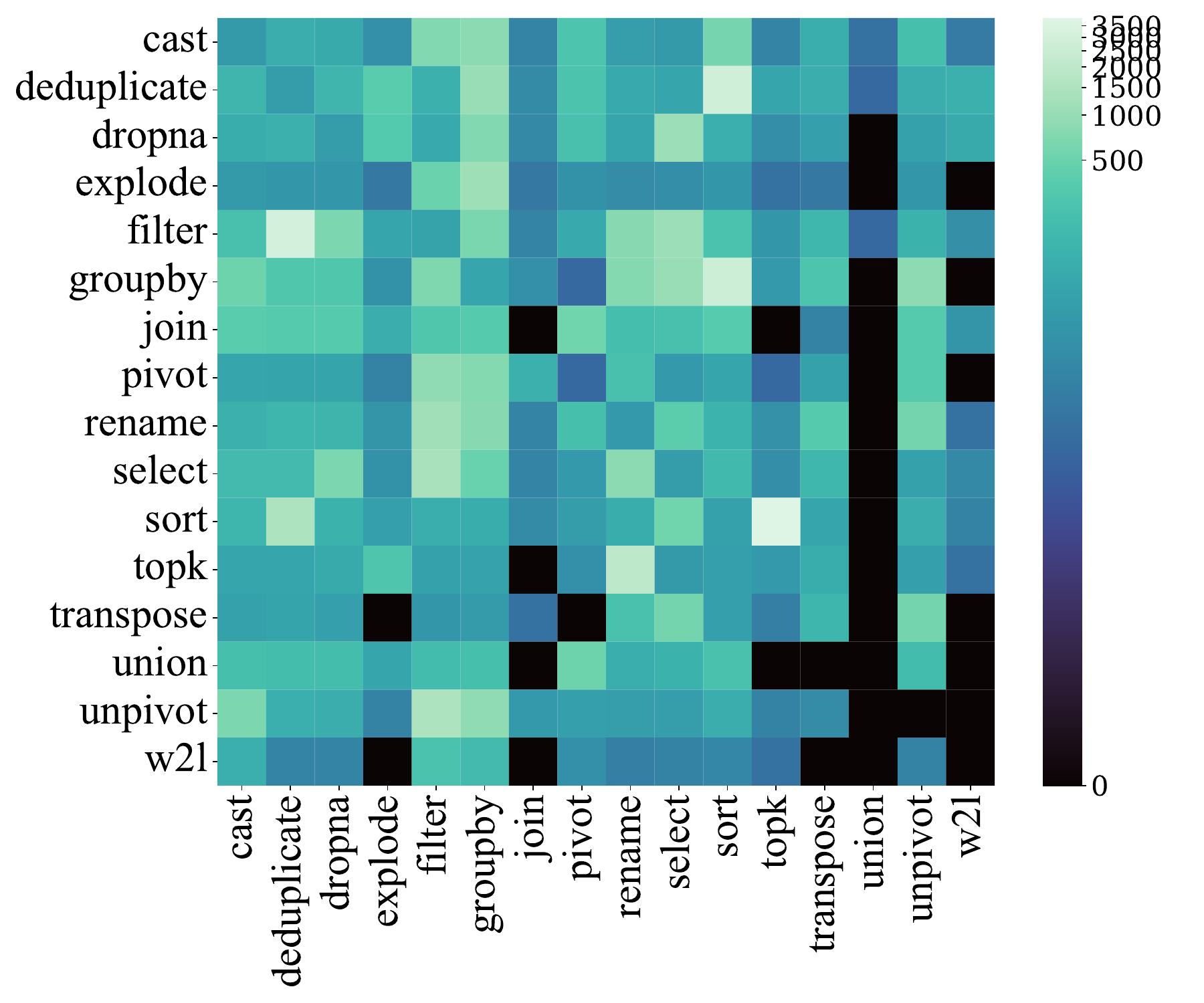}
  \end{minipage}
  \caption{Left: pipeline length distribution over three difficulty levels. Middle: instruction length distribution by token frequency. Right: operation transition matrix. The abbreviation ``w2l'' stands for the wide-to-long operator.}
  \label{fig:op-structural-charts}
\end{figure*}

\vpara{Compositional Complexity}
Fig.~\ref{fig:op-structural-charts}~(left) illustrates the distribution of transformation chain lengths. The benchmark is centered on non-trivial sequences, with 49.07\% of instances containing 4 to 6 operations and a significant tail of 18.57\% exceeding 7 steps. As shown in Tab.~\ref{tab:benchmark-stats}, the average chain length of 4.24 is substantially higher than prior pipeline synthesis benchmarks like Auto-Tables~\cite{auto-tables} (1.11) and comparable to Auto-pipeline~\cite{auto-pipeline} (4.10), but over a dataset that is orders of magnitude larger and driven by natural language. This distribution accurately reflects the compositional nature of real-world tasks. The core challenge here is not merely length, but the necessity for models to manage long-range dependencies and schema propagation, where the validity of an operation at step $t$ depends critically on the schema produced at step $t-1$.

\begin{table}[tbp]
    \centering
    \captionsetup{type=table}
    \captionof{table}{Statistics of the \smodel.}
    \label{tab:benchmark-stats}
    \begin{tabularx}{\linewidth}{@{}>{\raggedright\arraybackslash}p{0.5\linewidth} >{\centering\arraybackslash}p{0.49\linewidth}@{}}
        \toprule
        \textbf{Statistics} & \textbf{Number} \\
        \midrule
        \textbf{Total Tasks} & \textbf{17,168 (100\%)} \\
        \quad Single Tab. & 11,327 (66.0\%) \\
        \quad Multi-Tab. & 5,841 (34.0\%) \\
        \quad Train / Dev / Test  & 14,388 / 1,387 / 1,393 \\    
        \midrule
        \textbf{Input Table} & \\
        \quad Avg. Columns Per Tab. & 6.7 \\
        \quad Avg. Row Count & 134.2 \\
        \quad Num / Cat / Mixed & 44.3\% / 42.5\% / 13.2\% \\
        \midrule
        \textbf{Chain Complexity} & \\
        \quad Easy ($\leq$ 3 ops) & 32.36\% \\
        \quad Medium (4$\sim$6 ops) & 49.07\% \\
        \quad Hard ($\geq$7 ops) & 18.57\% \\
        \quad Avg. Chain Length & 4.24 \\
        \midrule
        \textbf{Instructions} & \\
        \quad Avg. Characters  & 192.3 \\
        \quad Avg. Tokens  & 100.5 \\
        \bottomrule
    \end{tabularx}
\end{table}

\begin{figure}[tbp]
    \centering
    \includegraphics[width=0.5\textwidth]{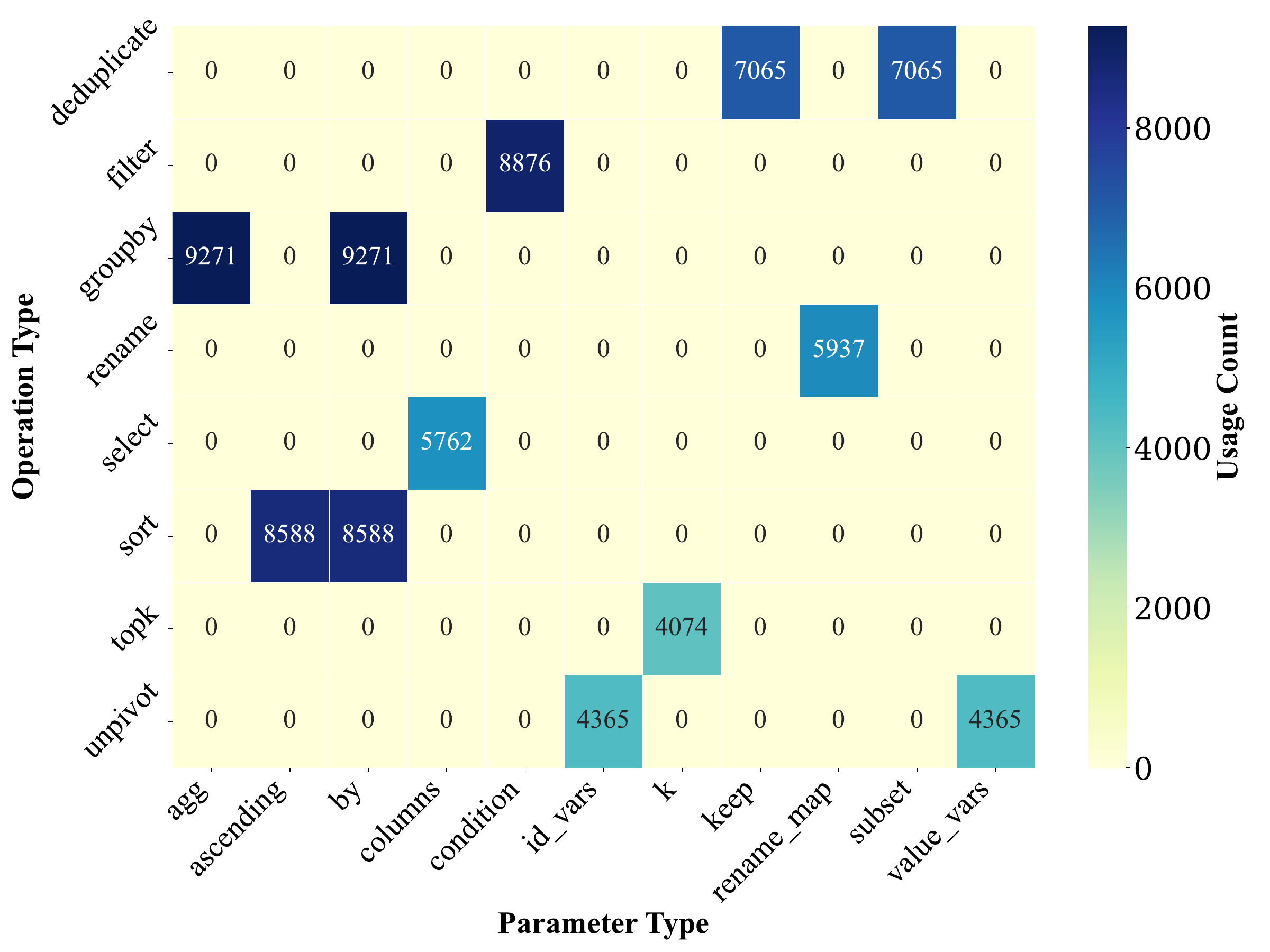}
    \caption{Heatmap of parameter usage across different operations. Darker colors indicate higher parameter complexity.}
    \label{fig:parameter_heatmap}
\end{figure}

\vpara{Parameter Complexity}
Beyond program depth, as illustrated in Fig.~\ref{fig:parameter_heatmap}, the parameter complexity varies significantly. While operations like \texttt{rename} and \texttt{select} are simple, structural operations like \texttt{join}, \texttt{pivot}, and \texttt{groupby} exhibit high parameter complexity. This presents a critical challenge, as these operations require models to generate structured arguments, not just single values. For instance, a \texttt{groupby} requires identifying \textit{both} the grouping keys (`by=...`) \textit{and} a dictionary of aggregation functions (`agg=...`). A \texttt{join} requires specifying join keys (`on=...`) and the join type (`how=...`). This transforms the task from simple sequence generation into a hierarchical prediction problem, requiring the model to deeply understand table semantics (e.g., key vs. value columns) to correctly populate these complex, structured parameters.

\subsection{Operational and Structural Diversity}
\begin{figure}[tbp]
    \centering
    \includegraphics[width=0.38\textwidth]{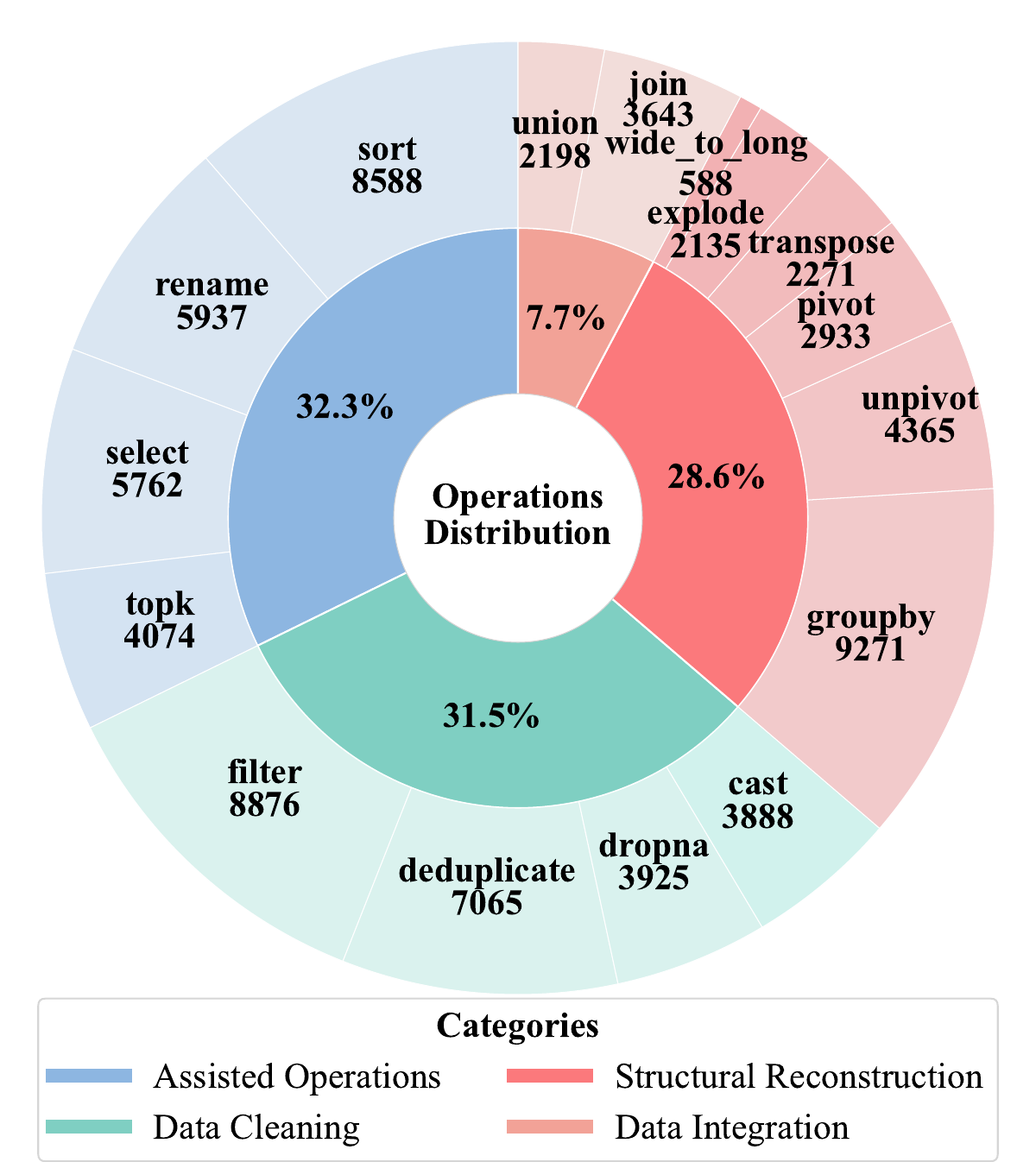}
    \caption{Operation distribution.}
    \label{fig:operation-distribution}
\end{figure}

As shown in Fig.~\ref{fig:operation-distribution}, \model supports 16 operations with distinct frequencies, capturing the breadth of data preparation requirements in real-world pipelines. 
Aggregation, reshaping, and integration together account for 48.6\% of operator usage, while selection and ordering operations provide complementary functionality.
This distribution effectively reflects empirically observed patterns in data science practice. Fig.~\ref{fig:op-structural-charts} (right) shows the operator transition graph among the most frequently occurring operations, revealing diverse and non-linear patterns among common operations. The dense connectivity and heterogeneous edge weights underscore the rich compositional patterns present in multi-step pipelines, necessitating sophisticated reasoning capabilities for models to successfully predict operation sequences that maintain schema compatibility and semantic coherence. 

\begin{figure}[tbp]
    \centering
    \includegraphics[width=1.0\linewidth]{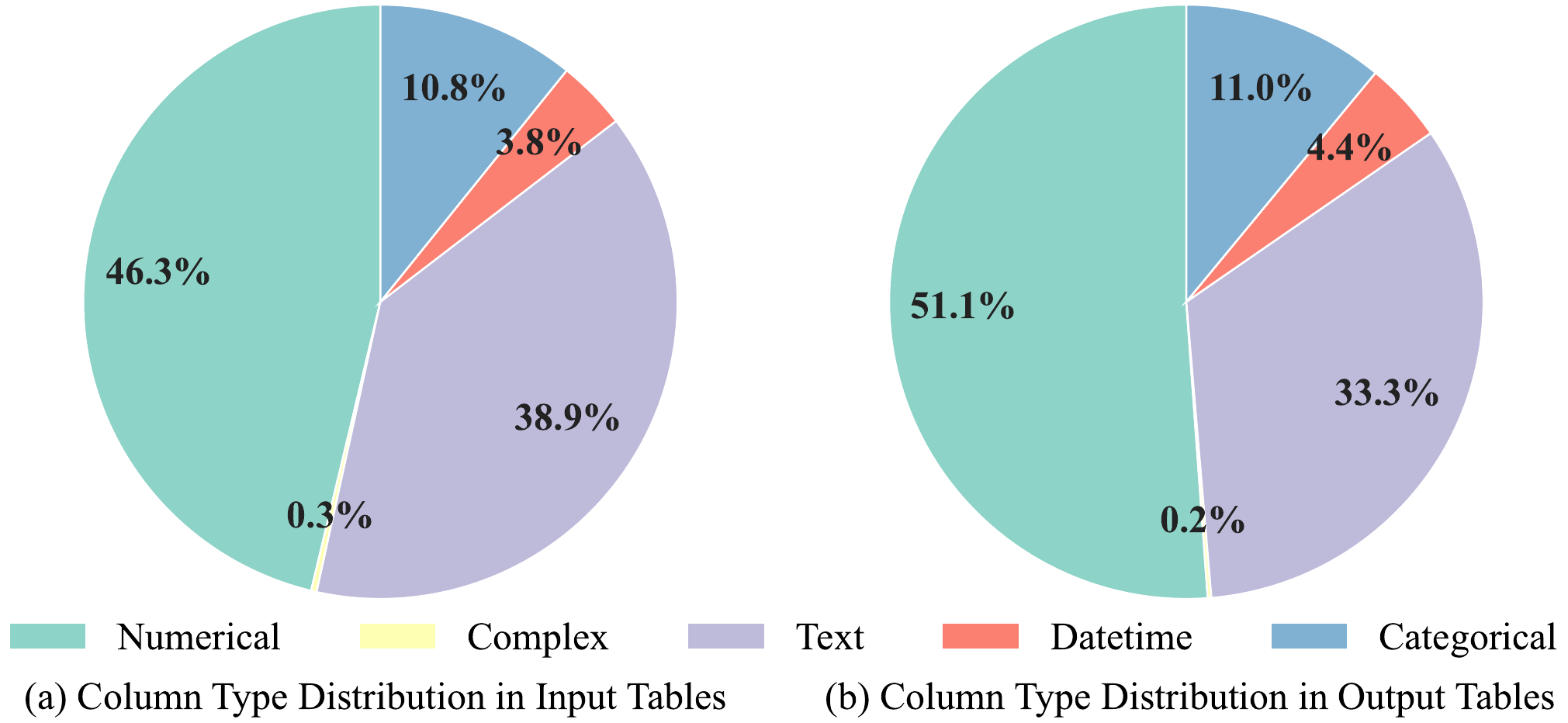}
    \caption{Column type distributions in input vs. output tables.}
    \label{fig:column_type_distribution}
\end{figure}

To further examine the diversity of the table, Fig.~\ref{fig:column_type_distribution} reports the distribution of column types in input versus output tables. Notably, output tables exhibit an increased proportion of textual columns and reduced numeric fields, reflecting structural transformations such as pivoting or aggregations that change schema layouts. These statistics collectively ensure that \model presents operational and structural diversity, challenging models across both symbolic planning and schema reasoning.

\subsection{Instruction Characteristics}
Fig.~\ref{fig:op-structural-charts}~(mid) illustrates the length distribution of instructions in \smodel, with an average of 107.4 tokens. The distributions are right-skewed and bimodal, reflecting both concise directives and longer, context-rich descriptions. 
Instruction complexity correlates strongly with transformation chain length, suggesting that linguistically complex prompts often entail more compositional operations. To quantify lexical and semantic diversity, we compute standard generation diversity metrics, including Distinct-n and Self-BLEU:
(1) \textbf{Distinct-n}~\citep{li2015diversity}: Measures the proportion of unique $n$-grams in the instruction corpus. Higher values indicate richer lexical variety. We report Distinct-1 and Distinct-2.
(2) \textbf{Self-BLEU}~\citep{zhu2018texygen}: Measures overlap between an instruction and the rest of the corpus. Lower Self-BLEU implies lower redundancy and more diverse phrasing.
Tab.~\ref{tab:diversity-metrics} summarizes the results. Compared to other instruction-driven datasets such as Spider and NL2Formula, \model exhibits significantly higher lexical diversity and lower redundancy, reflecting its open-ended, LLM-generated language design. 
These results confirm that \model instructions exhibit greater lexical richness and structural variation, which helps benchmark model generalization to diverse user intents and phrasing styles. Importantly, while the instructions are synthesized for control and consistency, they are grounded in thousands of noisy real-world tables, ensuring models must interpret instructions grounded in noisy and diverse schemas.
\begin{table}[tbp]
\centering
\caption{Lexical diversity comparison across datasets.}
\begin{tabular}{lccc}
\toprule
\textbf{Dataset} & \textbf{Distinct-1} & \textbf{Distinct-2} & \textbf{Self-BLEU-4} \\
\midrule
Spider~\cite{yu2018spider}         & 0.39 & 0.62 & 0.81 \\
NL2Formula~\cite{zhao2024nl2formula} & 0.42 & 0.68 & 0.77 \\
\textbf{\model (Ours)}             & \textbf{0.58} & \textbf{0.74} & \textbf{0.61} \\
\bottomrule
\end{tabular}
\label{tab:diversity-metrics}
\end{table}

\begin{figure}[tbp]
    \centering
    \includegraphics[width=1.0\linewidth]{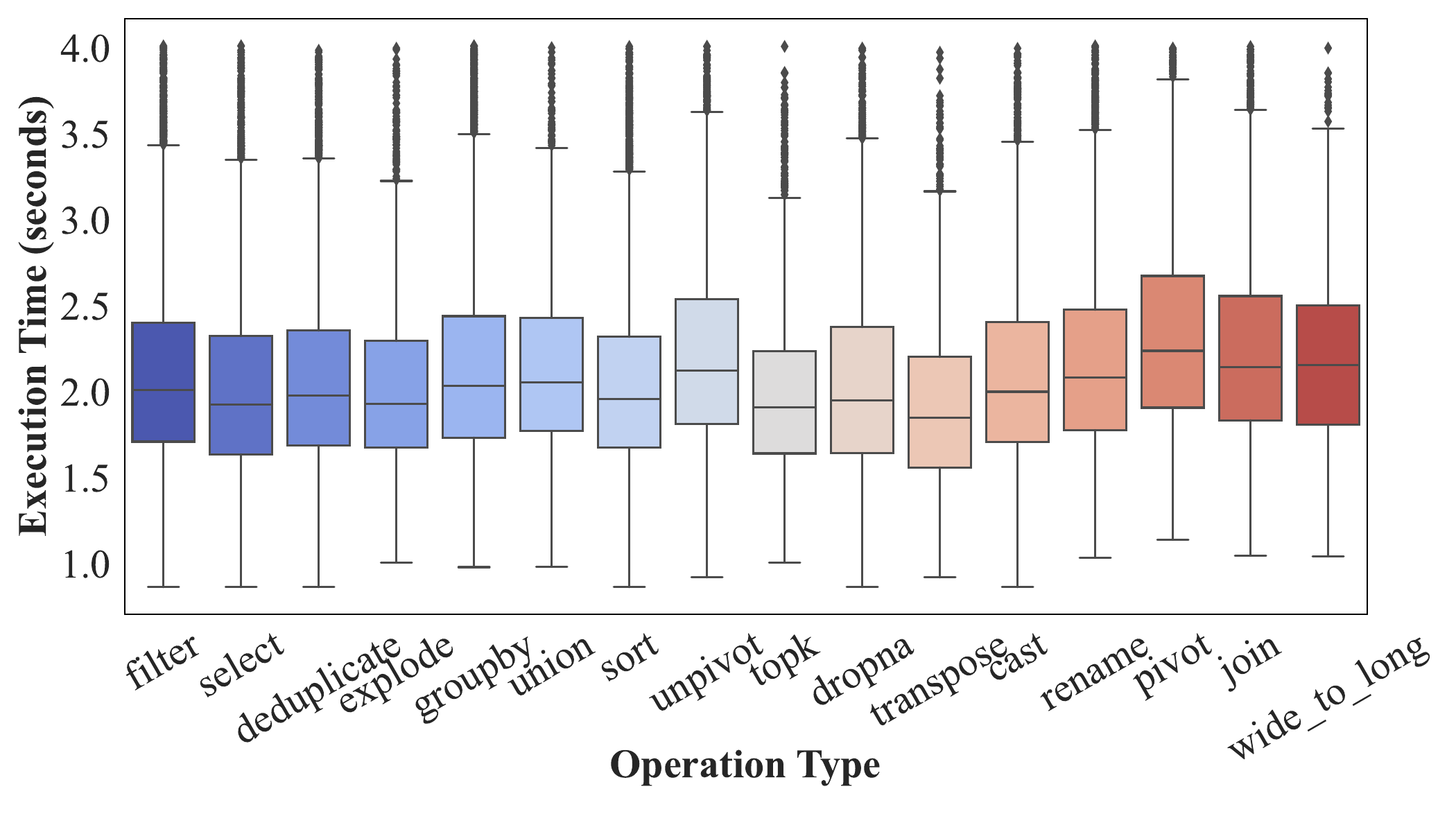}
    \caption{Distribution of per-operation execution times across all tasks, with the top 5\% of outliers removed} 
    \label{fig:op-vs-time}
\end{figure}


\subsection{Operator-Level Execution Characteristics} 
To quantify the computational costs of different operations, we analyze the per-operation execution time distribution across all tasks. Fig.~\ref{fig:op-vs-time} shows a box plot of execution times for each operation type, excluding the top 5\% of outliers. The analysis reveals a clear computational hierarchy: data-intensive operations such as \texttt{unpivot} and \texttt{union} exhibit both higher median execution times, while lightweight operations like \texttt{select} and \texttt{dropna} typically complete within 2 seconds. This computational heterogeneity highlights that operation choice and ordering can markedly affect total execution time, motivating operation-aware scheduling and synthesis.

\section{Pipeline-Agent}
\label{sec:pipeline-agent}

The \paradigm task presents two fundamental challenges that standard prompting methods are ill-equipped to handle: multi-step compositional logic and semantic parameter grounding. First, models must plan a correct, order-dependent sequence of operations. Second, they must correctly ground parameters (e.g., column names, aggregation functions) to a table schema that is dynamically evolving with each transformation step. 

Standard Tool Calling APIs~\cite{qin2024tool} typically execute one-shot instructions, lacking the dynamic state maintenance required for sequential transformations. Plan-and-Solve~\cite{wang2023plan}, which separates the planning phase from the execution phase, cannot easily leverage intermediate execution feedback to correct or adapt the plan. Similarly, while Chain-of-Tables~\cite{wang2024chain} focuses on static table reasoning, it lacks robust support for dynamic state tracking and adaptive tool chaining, making it less suitable for the iterative transformation tasks defined in our benchmark. To overcome these limitations, we introduce Pipeline-Agent, a unified framework designed to iteratively solve complex data preparation tasks by tightly coupling step-by-step reasoning with modular tool execution in a state-aware, feedback-driven interaction cycle. This agent's architecture is composed of three key components, which we elaborate on below.

\vpara{Cognitive Orchestrator} This component is the agent's planning and reasoning engine, specifically designed to tackle the compositional logic challenge. Specifically, we adopt a ReAct-style~\cite{yao2023react} approach, which is essential for handling the dynamic state changes inherent in \sparadigm. Unlike static Plan-and-Solve methods that commit to a full plan upfront, our orchestrator interleaves reasoning and execution. Each ``Thought'' is a structured reasoning step where the agent must (a) analyze the user's remaining intent, (b) inspect the current table state (the ``Observation''), and (c) formulate a single, executable ``Action''. This ``Action'' is a structured JSON call to a specific tool. After each action, the agent observes the outcome, such as the updated table schema or an execution error, which it as feedback for its next planning step. This allows the agent to dynamically refine its multi-step plan, correct mistakes, and handle complex dependencies based on observed table changes.

\vpara{Modular Toolsets} This component represents the agent's actionable capabilities, defined as a set of deterministic, modular tools. Each tool corresponds directly to one of the 16 core operations in our DSL (see Tab.~\ref{tab:dsl-operators}), such as \texttt{filter} or \texttt{groupby}. All tools adhere to a standard interface (e.g., \texttt{transform(df, **args)}) and accept structured JSON arguments. This design is the cornerstone of the agent's reliability: by forcing the LLM to generate only structured parameters rather than raw code, we mitigate the risks of direct code generation (e.g., LLM-hallucinated syntax, injection vulnerabilities, or use of non-existent \texttt{pandas} functions).  This abstraction ensures that the agent's actions are constrained, predictable, and robustly callable, preventing the LLM from generating syntactically incorrect or semantically unsafe code.

\vpara{Stateful Validator} This component acts as the feedback mechanism that grounds the agent's reasoning in executable reality. It serves two functions: execution and observation. First, it receives the ``Action'' (e.g., \texttt{\{'op': 'groupby', ...\}}) from the orchestrator, invokes the corresponding \textit{Modular Tool}, and executes it on the current table state $\mathbf{df}_t$. Second, it generates the ``Observation'' for the next reasoning step. This observation is not the full table data; it is a concise summary of the new state $\mathbf{df}_{t+1}$, including: (a) the updated table schema (column names and types), (b) a few sample rows (e.g., $df.head(5)$) to ground the LLM's understanding of the data values, and (c) a descriptive error message if the execution failed. This component effectively acts as the runtime-equivalent of the \textit{Schema Propagation Mechanism (SPM)} (Sec.~\ref{sec:operator-chain-construction}), ensuring that every step in the agent's plan is validated against a real, dynamically evolving schema, rather than a hypothesized one. In summary, these three components form a robust, self-correcting system. The agent's ``mind'' (Cognitive Orchestrator) is decoupled from its ``hands'' (Modular Toolsets), while the ``nerves'' (Stateful Validator) provide a high-fidelity feedback loop. It allows Pipeline-Agent to chain complex operations, correct errors in real-time, and dynamically adapt its plan based on intermediate results, thus overcoming the static, one-shot limitations of prior methods.

\FloatBarrier
\section{Experiments and Analysis}
\label{sec:experiments}

\subsection{Experimental Setup}
\label{sec:experimental_setup}

\begin{table*}[tbp]
    \centering
    \caption{Performance of baseline models on the \model test set across difficulty levels. We report EA, PV, and OA in percentages (\%). Best results per category are in \textbf{bold}.}
    \resizebox{\textwidth}{!}{
    \begin{tabular}{l cccc cccc cccc} 
    \toprule
    \multirow{2}{*}{\textbf{Model}} & \multicolumn{4}{c}{\textbf{Execution Accuracy (EA)}}              & \multicolumn{4}{c}{\textbf{Program Validity (PV)}}                & \multicolumn{4}{c}{\textbf{Operator Accuracy (OA)}}                \\ 
    \cmidrule(l){2-13}
                                    & Easy           & Medium         & Hard           & Overall        & Easy           & Medium         & Hard           & Overall        & Easy           & Medium         & Hard           & Overall         \\ 
    \midrule
    \multicolumn{13}{l}{\textbf{Zero-shot LLMs (non-reasoning)}}                                                                                                                                                                                 \\
    GPT-4o-mini                     & 75.17          & 59.55            & 49.79       & 62.88          & 87.92          & 73.83          & 62.34          & 76.38          & 67.08          & 70.44          & 69.6          & 69.22           \\
    GPT-4o                          & \textbf{89.03} & \textbf{79.06} & \textbf{72.38}          & \textbf{71.00} & 89.04 & \textbf{79.07} & \textbf{72.38}          & \textbf{81.12} & 64.24          & \textbf{70.70} & \textbf{74.44}          & \textbf{69.27}  \\
    Gemini-2.5-Pro                  & 80.09          & 62.09          & 52.72          & 66.26          & \textbf{91.5}          & 77.65          & 67.78 & 80.40          & \textbf{68.75} & 64.54          & 67.72          & 66.95           \\
    DeepSeek-V3                     & 78.52          & 63.79          & 56.07          & 67.19          & 89.49         & 77.09          & 68.62          & 67.31          & 66.70          & 68.32         & 72.62 & 68.54           \\ 
    \midrule
    \multicolumn{13}{l}{\textbf{Zero-shot LLMs (reasoning)}}                                                                                                                                                                                  \\
    GPT-o3-mini                         & \textbf{82.10} & 68.88  & \textbf{67.36} & \textbf{72.86} & 92.39          & 81.47 & 79.92 & 84.70 & \textbf{67.86} & \textbf{70.51} & \textbf{72} & \textbf{69.91}  \\
    GPT-o4-mini                     & 81.87          & \textbf{69.17}          & 64.02          & 72.36         & \textbf{94.85} & \textbf{83.31}          & \textbf{82.01}          & \textbf{86.79}          & 64.47          & 68.76          & 68.61          & 67.36           \\
    DeepSeek-R1                     & 77.18           & 58.84          & 41.84          & 61.81          & 89.49          & 72.70          & 55.23          & 75.09          & 67.45          & 68.06          & 67.39           & 67.75           \\ 
    \midrule
    \multicolumn{13}{l}{\textbf{Fine-tuned LLMs}}                                                                                                                                                                                            \\
    Qwen2.5-Coder-1.5B              & 67.79          & 59.41          & 50.63          & 60.59          & 77.85          & 70.58          & 64.02          & 71.79          & 67.11          & 70.33          & 66.55          & 68.65           \\
    Qwen2.5-Coder-3B                & 83.67          & 69.73          & \textbf{68.62}          & 74.01          & \textbf{91.05}          & 80.91          & \textbf{82.01}          & 84.35          & 82.03          & 81.76          & \textbf{81.93}          & 81.87           \\
    Qwen2.5-Coder-7B                & \textbf{84.12}	         & \textbf{69.87}	      & 68.20	       & \textbf{74.15}	        & 91.5	       & \textbf{82.18} 	        & 81.59	         & \textbf{85.07}	      & \textbf{82.96}	       & \textbf{82.96}	        & 81.59	       & \textbf{82.53}          \\   
    \bottomrule
    \end{tabular}}
\label{tab:baseline-performance}
\end{table*}

We conduct a comprehensive evaluation of various models on the \model benchmark to assess their capabilities in tackling the \paradigm task. Our experimental setup is detailed below.

\vpara{LLM Baselines}
To establish a comprehensive performance baseline on \smodel, we evaluate a diverse set of LLMs, encompassing both zero-shot inference capabilities of proprietary models and the performance of fine-tuned open-source models. These models represent the current state-of-the-art in code generation and natural language understanding: 
(1) \textbf{Zero-shot LLMs:} We utilize several leading API-based models, including GPT-4o~\cite{openai2023gpt}, GPT-4o-mini~\cite{openai2023gpt}, Gemini-2.5-Pro~\cite{reid2024gemini}, and DeepSeek-V3~\cite{liu2024deepseek}. These models are prompted with the task instruction and table schema without any task-specific fine-tuning.
(2) \textbf{Fine-tuned LLMs:} We fine-tune a series of strong open-source code generation models: Qwen2.5-Coder-1.5B~\cite{hui2024qwen2}, Qwen2.5-Coder-3B~\cite{hui2024qwen2}, and Qwen2.5-Coder-7B~\cite{hui2024qwen2}. These models are trained on the \model training set, which is derived from the \model benchmark, to adapt them specifically to the \paradigm task.

\vpara{Structured Generation Approaches}
We evaluate three distinct target output formalisms for generating executable data pipelines to understand their efficacy in representing and synthesizing complex transformations:
(1) \textbf{Text-to-Code (Pandas)}: Direct generation of executable Pandas code.
(2) \textbf{Text-to-SQL}: Generation of SQL statements, assessing the adaptability of SQL-centric approaches to broader data preparation tasks not typically addressed by SQL.
(3) \textbf{Text-to-Pipeline}: Our primary approach, where models generate operation sequences in our DSL, subsequently compiled to Pandas. The DSL is architected for modular planning, type safety through enforcement, and schema-aware validation, aiming for a more robust generation pathway.

\vpara{Planning and Agent-based Approaches}
To assess the capabilities of more sophisticated reasoning strategies for multi-step pipelines, we evaluate several planning and agent-based paradigms:
(1) \textbf{Tool Calling API~\cite{qin2024tool}}: LLMs are instructed to generate the full multi-step execution plan or program for a \model task in a single pass, simulating a direct tool-use scenario.
(2) \textbf{Plan-and-Solve~\cite{wang2023plan}}: This approach first generates a high-level operation plan~(e.g., a sequence of operations), then synthesizes the executable program based on that plan.
(3) \textbf{Chain-of-Tables~\cite{wang2024chain}}: This strategy involves evolving and manipulating intermediate tabular states throughout the reasoning chain to guide the transformation process for tasks from \smodel.
(4) \textbf{Pipeline-Agent}: Our proposed agent that iteratively predicts an operation, executes it on the current table, and reflects on the result. By leveraging intermediate states, it enables context-aware planning and handles schema evolution in complex \model tasks.


\vpara{Implementation Details} 
For zero-shot LLM evaluations, we utilized consistent prompt templates. Each prompt included: 
(1) a clear definition of the \paradigm task.
(2) the input table (column names and data types). 
(3) 10 sample rows from the input table to provide context on data values, and 
(4) the natural language instruction. We used a temperature of 0.7 for evaluation purposes. 
For fine-tuned models, we performed supervised fine-tuning on the \model training set. Models were trained for 3 epochs using the AdamW~\cite{kingma2014adam} optimizer with a learning rate of 1e-3 and a batch size of 16. A learning rate scheduler of linear decay with warmup was employed. Early stopping was triggered based on loss on a dedicated validation split of \model to prevent overfitting. All fine-tuning experiments were conducted on a cluster of 4 NVIDIA 4090 (24GB) GPUs. 
%
%
We use the GPT-4o-mini as the default LLM backbone unless specified otherwise. For Text-to-Pandas, the DSL-to-Pandas compilation is rule-based and deterministic. For Text-to-SQL, the generated SQL is subsequently executed via the SQLite engine. We have provided partial comments for operator types that are not supported by SQL. More experimental details and prompts design can be found in App.~\ref{app:experiments-detail} and App.~\ref{app:prompt-design}.

\subsection{Evaluation Results}
\label{sec:evaluation-results}
\begin{table*}[tbp]
    \centering
    \caption{Evaluation results of agent methods on the \model test set. We report EA, PV, and OA in percentages (\%). Best results per category are in \textbf{bold}. OA is omitted for Chain-of-Tables as it does not produce atomic operation sequences.}
    \label{tab:agent-planning-comparison}
    \resizebox{\textwidth}{!}{
    \begin{tabular}{lcccccccccccc} 
    \toprule
    \multirow{2}{*}{\textbf{Model}} & \multicolumn{4}{c}{\textbf{Execution Accuracy (EA)}} & \multicolumn{4}{c}{\textbf{\textbf{Program Validity (PV)}}} & \multicolumn{4}{c}{\textbf{Operator Accuracy (OA)}}  \\ 
    \cmidrule(l){2-13}
                                    & Easy  & Medium & Hard  & Overall                     & Easy  & Medium & Hard  & Overall                            & Easy  & Medium & Hard  & Overall                     \\ 
    \midrule
    Tool Calling                    & \textbf{71.62} & 58.36  & 47.67 & 60.48                       & 86.48 & 66.53  & 58.13 & 71.07                              & 67.79 & 47.15  & 35.40  & 51.31                       \\
    Plan-and-Solve                & 63.69 & 43.19  & 30.23 & 47.40                        & 74.52 & 52.53  & 38.37 & 57.00                                 & 61.04 & 42.83  & 30.09 & 46.36                       \\
    Chain-of-Tables                 & 50.67 & 17.90   & 9.30   & 26.27                       & 81.76 & 74.32  & 79.07 & 77.39                              & -     & -      & -     & -                           \\
    \midrule
    \multicolumn{13}{l}{\textbf{Pipeline-Agent}}                                                                                                                                                                         \\
    ~ ~ - GPT-4o-mini               & 70.27 & 61.08  & 54.65 & 62.72                       & 88.51 & 76.26  & 67.44 & 78.41                              & 67.56 & 52.04  & 41.02 & 54.79                       \\
    ~ ~ - GPT-4o                    & \textbf{77.70}  & \textbf{78.21}  & \textbf{67.44} & \textbf{76.17}                       & 96.62 & \textbf{88.33}  & \textbf{82.56} & \textbf{89.82}                              & \textbf{78.04} & \textbf{72.32}  & \textbf{65.91} & \textbf{72.92}                       \\
    ~ ~ - Deepseek-V3               & 66.89 & 60.70   & 48.84 & 60.49                       & \textbf{91.21} & 78.99  & 70.93 & 81.26                              & 68.47 & 58.18  & 47.53 & 59.42                       \\
    \bottomrule
    \end{tabular}}
\end{table*}

\begin{table*}[tbp]
\centering
\footnotesize
\caption{Performance comparison of different target generation formalisms on the \model test set. We report EA, PV, and OA in percentages (\%). Best results per category are in \textbf{bold}.}
\label{tab:target-generation-comparison}
\resizebox{\textwidth}{!}{
\begin{tabular}{lcccccccccccc} 
\toprule
\multirow{2}{*}{\textbf{Model}} & \multicolumn{4}{c}{\textbf{Execution Accuracy (EA)}}              & \multicolumn{4}{c}{\textbf{Program Validity (PV)}}                & \multicolumn{4}{c}{\textbf{Operator Accuracy (OA)}}                \\ 
\cmidrule(l){2-13}
& Easy & Medium & Hard  & Overall                  & Easy  & Medium & Hard  & Overall                   & Easy  & Medium & Hard & Overall                      \\ 
\midrule
Text-to-SQL                                & 10.81    & 2.59   & 0     & 3.05                     & \textbf{93.92} & 67.31  & 55.81 & 73.31                     & -     & -      & -    & -                            \\
Text-to-Code                               & 48.64    & 32.59  & 28.57 & 33.8                     & 70.27 & 57.77  & 53.74 & 58.45                     & -     & -      & -    & -                            \\
Text-to-Pipeline                           & \textbf{75.17}    & \textbf{59.55}  & \textbf{49.79} & \textbf{62.88}                    & 87.92 & \textbf{73.83}  & \textbf{62.34} & \textbf{76.38}                     & \textbf{67.08} & \textbf{70.44}  & \textbf{69.6} & \textbf{69.22}                        \\
\bottomrule
\end{tabular}}
\end{table*}

\vpara{Performance across Difficulty Levels}
Tab.~\ref{tab:baseline-performance} reports the performance of LLM baselines across Easy, Medium, and Hard tasks in \model. Zero-shot models with explicit reasoning prompts (e.g., GPT-o3-mini) outperform non-reasoning variants, achieving \textbf{72.86\%} EA vs. \textbf{71.00\%} (GPT-4o). GPT-4o (non-reasoning) performs well on Easy (\textbf{89.03\%}) and Medium (\textbf{79.06\%}) tasks, while showing a struggle with competitive Hard tasks (EA \textbf{71.00\%}). Fine-tuned models deliver the largest gains. Qwen2.5-Coder-7B achieves \textbf{74.15\%} overall EA and \textbf{68.20\%} on Hard tasks—outperforms leading closed-source LLMs~(Deepseek and GPT series) despite having fewer parameters. It also attains top PV (\textbf{85.07\%}) and OA (\textbf{82.53\%}), highlighting the quality of our supervised data.
Across all models, performance consistently declines from Easy to Hard tasks, with a sharper drop in EA than PV. This suggests that while many models produce syntactically valid outputs (high PV), ensuring execution correctness in complex, multi-step settings remains a core challenge. For instance, Qwen2.5-Coder-3B’s EA drops from \textbf{83.67\%} (Easy) to \textbf{68.62\%} (Hard), reflecting the compositional difficulty inherent.

\vpara{Impact of Structured Target Generation}
Our analysis confirms a core hypothesis of this work: targeting our DSL vastly outperforms direct code generation. 
%
%
%
As shown in Tab.~\ref{tab:target-generation-comparison}, it achieves an overall EA of \textbf{62.88\%}, exceeding Text-to-Code (Pandas, \textbf{33.80\%}) by \textbf{+29.08} points and Text-to-SQL (\textbf{3.05\%}) by a wide margin.
This advantage holds across difficulty levels, particularly on Easy (\textbf{75.17\%}) and Medium (\textbf{59.55\%}) tasks. Although Text-to-SQL yields high PV (\textbf{73.31\%}), its low EA indicates poor semantic grounding in complex, multi-step tasks beyond standard SQL patterns.
Text-to-Pipeline also achieves the highest OA (\textbf{69.22\%}), reflecting stronger structural fidelity. OA is not reported for Text-to-Code and Text-to-SQL since their outputs do not use atomic operations as our DSL. These results highlight the DSL’s effectiveness in enabling more accurate planning and execution for compositional data preparation tasks.

\vpara{Efficacy of Planning and Agent-based Approaches}
Structured planning with agent-based methods significantly improves multi-step reasoning. As shown in Tab.~\ref{tab:agent-planning-comparison}, our proposed Pipeline-Agent (GPT-4o) achieves the highest overall EA of \textbf{76.17\%}, outperforming Tool Calling API (\textbf{60.48\%}) and Plan-and-Solve (\textbf{47.40\%}) by large margins. Even with the weaker GPT-4o-mini backbone, Pipeline-Agent still outperforms both baselines (\textbf{62.72\%} EA), confirming its robustness.
Chain-of-Tables, while exhibiting strong PV on Hard tasks (\textbf{79.07\%}), suffers from poor EA (\textbf{26.27\%}), likely due to unstable intermediate manipulations. Notably, upgrading the Pipeline-Agent's backbone from GPT-4o-mini to GPT-4o yields a substantial \textbf{+13.45} point EA gain, demonstrating its ability to leverage stronger models effectively. Across difficulty levels, Pipeline-Agent (GPT-4o) maintains high accuracy: \textbf{77.70\%} (Easy), \textbf{78.21\%} (Medium), and \textbf{67.44\%} (Hard).
Its high OA (\textbf{72.92\%} with GPT-4o, \textbf{54.79\%} with GPT-4o-mini) further highlights its strength in structuring valid transformation sequences. Since Chain-of-Tables produces non-symbolic intermediate states, OA is not reported for this method. These results underscore that agent-based strategies with explicit planning are key to tackling the compositional challenges, especially when paired with capable LLMs.

\subsection{Error Analysis and Case Study}
\label{sec:error_analysis}
\begin{figure*}[tbh]
    \centering
    \includegraphics[width=1.0\linewidth]{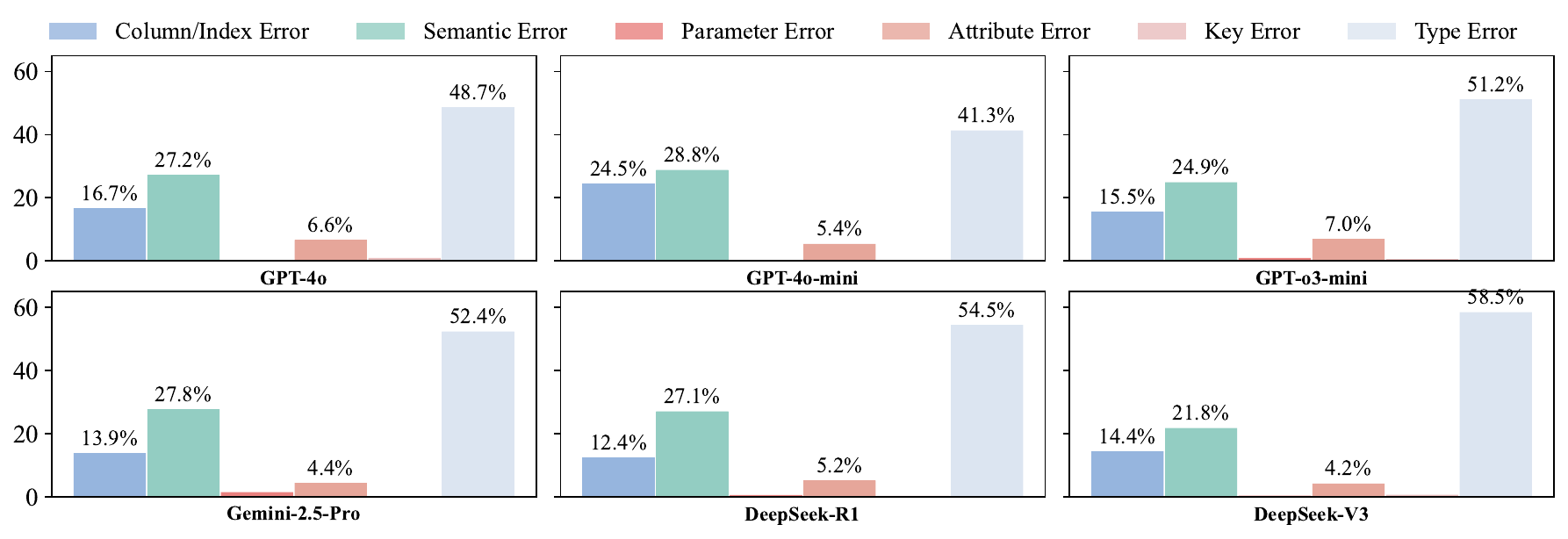}    
    \caption{Distribution of error types across six large language models.}
   \label{fig:error-type-distribution}
\end{figure*}
\begin{figure}[thb]
    \centering
    \includegraphics[width=1.0\linewidth]{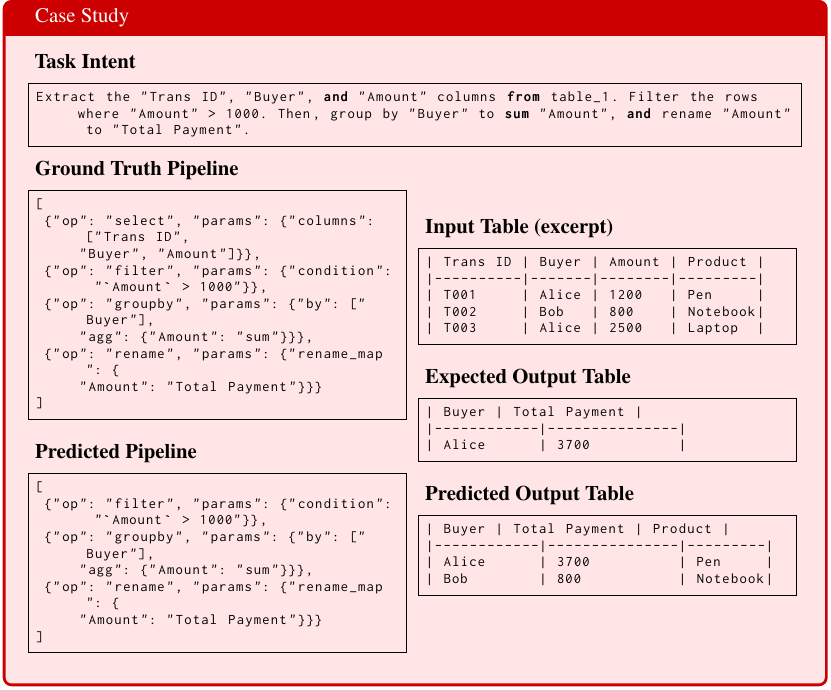}
    \caption{Case study with zero-shot prompting.}
    \label{fig:case_study}
\end{figure}

\vpara{Error Analysis} We conduct a comparative error analysis across six leading LLMs, as illustrated in Fig.~\ref{fig:error-type-distribution}, to understand common failure modes. We summarize key failure modes below, which we find map directly to our two identified core challenges:
\noindent (1) \textbf{Type errors} consistently account for the largest proportion across all models, ranging from 41.3\% (GPT-4o-mini) to 54.5\% (DeepSeek-R1). These errors typically stem from applying transformations to columns with incompatible data types, such as aggregating over non-numeric fields or sorting categorical columns without conversion. Models with weaker schema-tracking capabilities (e.g., DeepSeek-R1/V3) exhibit higher type error rates. This is a primary manifestation of failed semantic parameter grounding.
\noindent (2) \textbf{Semantic errors} constitute the second largest category, with models like GPT-4o and Gemini-2.5-Pro showing around 27\% semantic failures. These reflect incorrect or missing operations, broken logic chains, or hallucinated steps that deviate from the task intent. The relatively lower semantic error rates of DeepSeek models may indicate a conservative generation style, albeit at the cost of lower execution accuracy. This directly reflects a failure in multi-step compositional logic.
\noindent (3) \textbf{Column/index errors} are more prominent in GPT-o3/4-mini and Gemini-2.5-Pro, often resulting from misaligned references due to schema evolution (e.g., renaming or selection). This suggests limitations in maintaining coherent schema state across steps. This is another key failure mode of semantic parameter grounding under dynamic schema changes.
\noindent (4) \textbf{Attribute errors} (e.g., wrong aggregation function or sorting order) appear in 3\%--7\% of cases across models, indicating shallow mapping between instruction semantics and operator parameters. Overall, these results highlight the \textit{diverse failure modes} of different LLMs in \sparadigm, and emphasize the importance of schema tracking, operator grounding, and long-range reasoning in achieving robust program synthesis. This also represents a fine-grained failure in semantic grounding.

\vpara{Case Study}
We examine a failure case from zero-shot prompting, shown in Fig.~\ref{fig:case_study}, where the model is instructed to perform a series of operations: extract relevant columns (\texttt{select}), filter rows by value (\texttt{filter}), aggregate results (\texttt{groupby}), and finally rename a column (\texttt{rename}). The user instruction states:
\begin{quote}
\small\emph{``Extract the `Transaction ID', `Buyer Name', and `Amount Paid' columns from table\_1. Filter the rows where `Amount Paid' is greater than 1000. Then, group the data by `Buyer Name' and calculate the total sum of `Amount Paid' for each buyer. Finally, rename `Amount Paid' to `Total Payment'.''}
\end{quote}
The ground truth program includes all four operations in the specified order. However, the model-generated pipeline omits the initial \texttt{select} step and begins directly with \texttt{filter}, followed by \texttt{groupby} and \texttt{rename}. This mistake illustrates the critical failure modes as we mentioned before: (1) Failure in multi-step compositional Logic. The model fails to generate the correct, order-dependent sequence, omitting the initial \texttt{select} step entirely. It fails to reason that this column-scoping operation is a necessary compositional precursor, forcing downstream operators to process an incorrect (wider) schema and deviating from the required plan. (2) Failure in semantic grounding. While the model correctly parameterizes the operations it generates (e.g., `by=[``Buyer'']`), it fails to ground the entire instruction—specifically, it ignores the ``Extract...'' directive. This failure to map the full semantic intent of the NL to a complete symbolic program is a fundamental grounding failure.
This case study demonstrates that even when a model avoids simple parameter errors, it can still fail compositionally and fail to ground the user's complete intent, thus validating the challenges of \smodel.

\section{Conclusion}
\label{sec:conclusion}

In this work, we defined the \paradigm task and constructed \smodel, a large-scale benchmark designed to support systematic evaluation. Our experiments validated \smodel's role as a critical and difficult testbed. We demonstrated that the choice of representation is crucial, as direct code generation fails markedly (e.g., 33.8\% EA for Pandas), while our symbolic DSL provides a more robust target. Furthermore, our in-depth error analysis pointed to multi-step compositional logic and semantic grounding as the key bottlenecks. While our proposed \textit{Pipeline-Agent} achieves the SOTA, its significant remaining performance gap underscores the depth of the challenge. Future work could focus on developing agents that better track dynamic table states throughout the pipeline. For instance, reinforcement learning could be employed to train a policy, using rich execution errors as negative reward signals. \model provides the community with a critical benchmark for developing the next generation of data preparation agents.

\begin{acks}
 This work was supported by the [...] Research Fund of [...] (Number [...]). Additional funding was provided by [...] and [...]. We also thank [...] for contributing [...].
\end{acks}

\clearpage

\bibliographystyle{ACM-Reference-Format}
\bibliography{refers}

\appendix

\section{Details of Experiments}
\label{app:experiments-detail}

\subsection{LLM Baselines}

\vpara{Zero-shot LLMs}
\label{app:Zero-shot-LLMs}
To ensure consistent evaluation across models, we adopt a unified prompt format detailed in Section~\ref{zero-shot-prompt}. This schema-aware and instruction-driven template provides clear guidance on operation selection and output structure. During inference, all models are run with a temperature of 0.7 and a maximum output length of 4000 tokens. For models without a \texttt{temperature} parameter (e.g., \texttt{o3-mini}, \texttt{o4-mini}), this setting is omitted.

\vpara{Fine-tuned LLMs}
\label{app:fine-tuning}
To align large language models with data preparation tasks, we fine-tune several instruction-tuned variants of Qwen2.5-Coder, including 1.5B, 3B, and 7B models. Training is performed on four NVIDIA RTX~4090 GPUs using \texttt{torchrun} with mixed precision (bf16). The batch size is set to 2 for the 1.5B model and 1 for the 3B and 7B models. Each model is trained for three epochs with a sequence length of 4096. 


\subsection{Structured Generation Approaches}

\vpara{Text-to-Code (Pandas)}
\label{app:text-to-pandas}
The Text-to-Pandas module converts natural language instructions into executable \texttt{pandas} code. It employs structured, context-aware prompts that help the model identify operation types (e.g., \texttt{filter}, \texttt{groupby}, \texttt{pivot}) and generate semantically correct, executable code based on the table schema. This design lowers the barrier to complex transformations and supports reproducible evaluation. Detailed templates are provided in Section~\ref{text-to-pandas-prompt}.

\vpara{Text-to-SQL}
\label{app:text-to-sql}
For the Text-to-SQL setting, we simulate a realistic database environment using SQLite, where all benchmark tables are stored as normalized relational tables with explicit schemas and foreign key relationships to support joins. This setup ensures that model-generated SQL queries are executable and verifiable. Structured prompt templates (Section~\ref{text-to-sql-prompt}) pair natural language instructions with schema descriptions to promote consistent and accurate SQL generation.

\subsection{Planning and Agent-based Methods}
\label{app:agent-methods}


\vpara{Tool and Environment Setup}
Each data preparation operation is implemented as a callable tool (Section~\ref{tool_definition}). Agents invoke these tools via LangChain’s \texttt{tool calling} API within a controlled Python environment that provides access to input tables, schemas, and intermediate states.

\vpara{Tool Calling Agent}
This baseline directly maps user instructions to tool invocations using LLM reasoning. The input includes a table preview, natural language instruction, and standardized tool schema documentation. We evaluate GPT-4o, GPT-4o-mini, and DeepSeek-V3 under this setting.

\vpara{Plan-and-Solve Agent}
Following a two-phase paradigm, this agent first generates a high-level transformation plan (e.g., \texttt{filter → groupby → sort}) and then executes each operation sequentially through tool calls. This decouples reasoning from execution but lacks feedback integration.

\vpara{Chain-of-Tables Agent}
Inspired by~\cite{wang2024chain}, this agent explicitly maintains intermediate tabular states, enabling stepwise manipulation and reasoning across evolving tables. It adapts plan–execute–react loop to operate over our DSL-based transformation layer instead of SQL.

\vpara{Pipeline-Agent}
We propose Pipeline-Agent, a unified framework that couples reasoning and execution in a closed-loop cycle. At each step, the agent reasons about the current table state, selects the next operation, executes it through the corresponding tool, and incorporates feedback into subsequent reasoning. This tight integration of reasoning and execution enables robust handling of complex, multi-step transformations. We evaluate Pipeline-Agent with GPT-4o, GPT-4o-mini, and DeepSeek-V3 to analyze the impact of reasoning capacity on performance.

\subsection{Tool Definitions and Interaction}
\label{tool_definition}
Each tool in the Pipeline-Agent is a self-contained module that encapsulates a particular transformation logic. Tools expose a unified interface for execution and are compatible with structured reasoning inputs from LLMs. During the agent’s reasoning process, these tools are dynamically selected and applied, enabling seamless integration into the closed-loop pipeline. The toolset spans essential operations such as filtering, grouping, sorting, pivoting, and more, supporting a wide range of data preparation needs.

\begin{tcolorbox}[colback=lightgreen,colframe=darkgreen, breakable, title=Tools Definition, label=tools]
\begin{Verbatim}[breaklines=true]
class BaseOpInput(BaseModel): 
    table_names: str
    
class FilterInput(BaseOpInput): 
    condition: str
    
class SortInput(BaseOpInput): 
    by: List[str]; ascending: List[bool]
    
class PivotInput(BaseOpInput): 
    index: str 
    columns: str 
    values: str 
    aggfunc: str
    
class StackInput(BaseOpInput): 
    id_vars: List[str]
    value_vars: List[str]
    
class ExplodeInput(BaseOpInput): 
    column: str 
    split_comma: bool
    
class WideToLongInput(BaseOpInput): 
    subnames: List[str] 
    i: List[str] 
    j: str 
    sep: str 
    suffix: str
    
class UnionInput(BaseOpInput):
    left_table: str 
    right_table: str
    how: str
    
class JoinInput(BaseOpInput):
    left_table: str
    right_table: str
    left_on: str
    right_on: str
    how: str
    suffixes: List[str]
    
class TransposeInput(BaseOpInput): pass

class DropnaInput(BaseOpInput): 
    subset: List[str]; 
    how: str
    
class DeduplicateInput(BaseOpInput): 
    subset: Union[List[str], None] 
    keep: str
    
class TopKInput(BaseOpInput): 
    k: int
    
class SelectInput(BaseOpInput): 
    columns: List[str]
    
class CastInput(BaseOpInput): 
    column: str
    dtype: str
    
class RenameItem(BaseModel):
    old_name: str
    new_name: str
    
class RenameInput(BaseOpInput):
    rename_items: List[str]
    
class AggItem(BaseModel):
    column: str
    agg_func: str
    
class GroupByInput(BaseOpInput):
    by: List[str]
    aggregations: List[AggItem]
\end{Verbatim}
\end{tcolorbox}

\subsection{Difficulty Level Definition and Illustrative Examples}
\label{app:difficulty}

\vpara{Difficulty Level Definitions}
To simulate tasks of varying complexity, we classify operator chains into three difficulty levels based on chain length and the semantic nature of operations involved:
\begin{itemize}[leftmargin=*]
  \item \textbf{Easy} (1–3 steps):  
  These chains typically consist of atomic, table-local operations such as \texttt{filter}, \texttt{sort}, \texttt{select}, or \texttt{dropna}.  

  \item \textbf{Medium} (4–6 steps):  
  Chains in this category often involve combinations of aggregation and light integration, such as \texttt{groupby}-\texttt{agg}, \texttt{rename}, or simple \texttt{join} operations.  

  \item \textbf{Hard} (7–8 steps):  
  These chains incorporate multi-table joins, nested reshaping (e.g., \texttt{pivot} following \texttt{groupby}), and schema-evolving transformations requiring global reasoning.  
  %
\end{itemize}

This stratified scheme enables us to evaluate model performance across various compositional depths and reasoning challenges, while ensuring control over the distribution of task complexity in the benchmark.

\vpara{Illustrative Examples} There are three examples for each level of difficulty. Each example consists of a natural language instruction, the corresponding DSL operator sequence, the compiled code, and input/output tables. These examples demonstrate different reasoning demands, such as filtering, sorting and multi-table aggregation like union and joining with nested operations.

\begin{tcolorbox}[colback=lightgreen,colframe=darkgreen, breakable,title=Example (Easy)]
\begin{Verbatim}[breaklines=true]
{
  "instruction": "Sort the data in 'table_1' by 'Civil Liberties' and 'President' in ascending order to organize the entries accordingly.",
  "input_table": "input_E001.csv",
  "output_table": "output_E001.csv",
  "transformation_sequence": [
  "op": "sort", "params": {"by": ["Civil Liberties","President"],"ascending": [true,true]},}
  ],
  "gold_code": "df.sort_values(by=['Civil Liberties', 'President'], ascending=[True, True]))"
}
\end{Verbatim}
\end{tcolorbox}

\vpara{Input Table (E001)}
\begin{center}
\resizebox{\columnwidth}{!}{
\begin{tabular}{ccccc}
\toprule
\textbf{Year} & \textbf{Political Rights} & \textbf{Civil Liberties} & \textbf{Status} & \textbf{President} \\
\midrule
1972 & 6 & 6 & Not Free & Hamani Diori \\
1973 & 6 & 6 & Not Free & Hamani Diori \\
1974 & 7 & 6 & Not Free & Hamani Diori \\
1975 & 7 & 5 & Not Free & Seyni Kountché \\
1976 & 7 & 5 & Not Free & Seyni Kountché \\
1977 & 7 & 5 & Not Free & Seyni Kountché \\
\bottomrule
\end{tabular}
}
\end{center}

\vpara{Output Table (E001)}
\begin{center}
\resizebox{\columnwidth}{!}{
\begin{tabular}{lcccc}
\toprule
\textbf{Year} & \textbf{Political Rights} & \textbf{Civil Liberties} & \textbf{Status} & \textbf{President} \\
\midrule
1975 & 7 & 5 & Not Free & Seyni Kountché \\
1976 & 7 & 5 & Not Free & Seyni Kountché \\
1977 & 7 & 5 & Not Free & Seyni Kountché \\
1972 & 6 & 6 & Not Free & Hamani Diori \\
1973 & 6 & 6 & Not Free & Hamani Diori \\
1974 & 7 & 6 & Not Free & Hamani Diori \\
\bottomrule
\end{tabular}
}
\end{center}

\begin{tcolorbox}[colback=lightgreen,colframe=darkgreen, breakable,title=Example (Medium)]
\begin{Verbatim}[breaklines=true]
{
  "instruction": "Start by excluding the rows where the 'Year' is 2013. Then, remove duplicate rows in table_1, keeping the last occurrence for each duplicate. After that, group the resulting data by 'Name', computing the minimum of 'Number of Contestants' for each group. Finally, sort the grouped data by 'Number of Contestants' in ascending order.",
  "input_table": "input_M001.csv",
  "output_table": "output_M001.csv",
  "transformation_sequence": [
    {"op": "filter", "params": {"column": "Year", "condition": "!= 2013"}},
    {"op": "duplicate", "params": {"subset": null, "keep": "last"}},
    {"op": "groupby", "params": {"by": ["Name"], "agg": {"Number of Contestants": "min"}}},
    {"op": "sort", "params": {"by": ["Number of Contestants"], "ascending": [true]}}
  ],
  "gold_code": "df.query('Year != 2013')
    .drop_duplicates(keep='last')
    .groupby('Name', as_index=False)
    .agg({'Number of Contestants': 'min'})
    .sort_values(by='Number of Contestants', 
    ascending=True)"
}
\end{Verbatim}
\end{tcolorbox}

\vpara{Input Table (M001)}
\begin{center}
\footnotesize

\begin{tabularx}{\columnwidth}{X c c c}
\toprule
\textbf{Name} & \textbf{\makecell{Number of \\ Contestants}} & \textbf{\makecell{Number of \\ Approved}} & \textbf{Year} \\
\midrule
University of Chile & 253 & 125 & 2014 \\
Pontifical Catholic University of Chile & 202 & 118 & 2014 \\
University of Concepción & 108 & 46 & 2013 \\
University of Chile & 74 & 33 & 2015 \\
Pontifical Catholic University of Chile & 69 & 31 & 2015 \\
\bottomrule
\end{tabularx}
\end{center}

\vpara{Output Table (M001)}
\begin{center}
\resizebox{\columnwidth}{!}{
\begin{tabular}{lc}
\toprule
\textbf{Name} & \textbf{Number of Contestants} \\
\midrule
Pontifical Catholic University of Chile & 69 \\
University of Chile & 74 \\
\bottomrule
\end{tabular}}
\end{center}

\begin{tcolorbox}[colback=lightgreen,colframe=darkgreen, breakable,title=Example (Hard)]
\begin{Verbatim}[breaklines=true]
{
  "instruction": "Start by performing a right join between table_1 and table_2 on 'ship id' with suffixes '_left' and '_right'. Then, remove any rows with missing values across all columns. Next, explode the 'location' column by splitting its values at commas. Group the data by the 'type' column, counting occurrences of 'speed knots' and calculating the mean of 'ship id'. Sort the grouped data by 'speed knots' and 'ship id' in descending order. Deduplicate the results based on 'speed knots' and 'ship id', keeping the first occurrence. Select the top two entries from the sorted results. Finally, rename the columns to 'category' for 'type', 'velocity in nautical miles' for 'speed knots', and 'vessel identifier' for 'ship id'.",
  "input_table": [ 
    "input_H001_ships.csv",
    "input_H001_ship_missions.csv"
   ],
  "output_table": "output_H001.csv",
  "transformation_sequence": [
    { "op": "join", "params": { "on": "ship id", "how": "right", "suffixes": ["_left", "_right"] } },
    { "op": "dropna", "params": { "how": "all" } },
    { "op": "explode", "params": { "column": "location", "split_comma": true } },
    { "op": "groupby", "params": { "by": ["type"], "agg": { "speed knots": "count", "ship id": "mean" } } },
    { "op": "sort_values", "params": { "by": ["speed knots", "ship id"], "ascending": [false, false] } },
    { "op": "deduplicate", "params": { "subset": ["speed knots", "ship id"], "keep": "first" } },
    { "op": "head", "params": { "n": 2 } },
    { "op": "rename", "params": { "rename_map": { "ship id": "vessel identifier", "speed knots": "velocity in nautical miles", "type": "category" } } }
  ],
  "gold_code": "df = ( 
    table_1.merge(table_2, on='ship id', how='right', suffixes=('_left', '_right')) 
    .dropna(how='all') 
    .assign(location=lambda df: df['location'].str.split(',')) 
    .explode('location') 
    .groupby('type', as_index=False) 
    .agg({'speed knots': 'count', 'ship id': 'mean'}) 
    .sort_values(by=['speed knots', 'ship id'], ascending=[False, False]) 
    .drop_duplicates(subset=['speed knots', 'ship id'], keep='first') 
    .head(2) 
    .rename(columns={ 
        'ship id': 'vessel identifier', 
        'speed knots': 'velocity in nautical miles', 
        'type': 'category' 
    })"
}
\end{Verbatim}
\end{tcolorbox}


\vpara{Input Table 1: Ships (H001)}
\begin{center}
\resizebox{\columnwidth}{!}{
\begin{tabular}{ccccc}
\toprule
\textbf{ship id} & \textbf{name} & \textbf{type} & \textbf{nationality} & \textbf{tonnage} \\
\midrule
1 & Corbridge & Cargo ship & United Kingdom & 3687 \\
2 & Farringford & Battle ship & United States & 3146 \\
3 & Dromonby & Cargo ship & United Kingdom & 3627 \\
4 & Author & Cargo ship & United Kingdom & 3496 \\
5 & Trader & Battle ship & United Kingdom & 3608 \\
6 & Ariadne & Cargo ship & United States & 3035 \\
7 & Appam & Battle ship & United Kingdom & 7781 \\
8 & Clan McTavish & Cargo ship & United States & 5816 \\
\bottomrule
\end{tabular}
}
\end{center}

\vpara{Input Table 2: Ship\_missions (H001)}
\begin{center}
\setlength{\tabcolsep}{3pt}
\footnotesize

\begin{tabularx}{\columnwidth}{c c c c c c}
\toprule
\textbf{\makecell{mission \\ id}} & \textbf{\makecell{ship \\ id}} & \textbf{\makecell{launched \\ year}} & \textbf{location} & \textbf{\makecell{speed \\ (knots)}} & \textbf{fate} \\
\midrule
1 & 1 & 1930 & Germany & 25 & Decommissioned \\
2 & 2 & 1930 & Germany & 25 & Decommissioned \\
3 & 3 & 1930 & Helsinki, Finland & 23 & Lost \\
4 & 5 & 1916 & Norway & 16 & Retired \\
5 & 6 & 1931 & Uusikaupunki, Finland & 23 & Decommissioned \\
6 & 7 & 1931 & Uusikaupunki, Finland & 23 & Decommissioned \\
7 & 8 & 1932 & Turku, Finland & 23 & Lost \\
\bottomrule
\end{tabularx}
\end{center}

\vpara{Output Table (H001)}
\begin{center}
\begin{tabular}{ccc}
\toprule
\textbf{category} & \textbf{velocity in nautical miles} & \textbf{vessel identifier} \\
\midrule
Cargo ship & 7 & 4.875 \\
Battle ship & 4 & 5.25 \\
\bottomrule
\end{tabular}
\end{center}

\section{Prompt Design}
\label{app:prompt-design}

\subsection{Prompts for Data Synthesis}
\label{app:data-synthesis-prompts}

In this section, we detail the prompts employed in our data synthesis framework. As previously mentioned in Section \ref{sec:benchmark-design}, we primarily use LLMs for generating and refining instructions, as well as for verifying the consistency between natural language instructions and the operator chain in the DSL. The specifics of these prompts are as follows:
\vspace{0.5em}
\hypertarget{prompt-generation}{}
\noindent
\begin{tcolorbox}[colback=blue!10,colframe=blue!50!black, breakable, title={Prompt for Instruction Generation}, label=prompt-step1]
\begin{Verbatim}[breaklines=true]
You are a data preparation expert. I have some related input tables and a target table, where the target table is obtained by transforming the input tables. Based on the transformation relationship between them, please generate a clear natural language instruction that describes how to transform the input tables into the target table.

The transformation operations and their detailed parameters are as follows:  
{transform_chain_str}

Input Tables (First 10 Rows):  
{input_table_str}

Target Table (First 10 Rows):  
{target_table_str}

Please generate a clear and natural data preparation instruction in English. The instruction should explicitly describe the required transformation steps and clearly state the table names involved, without mentioning specific programming languages or function names. Use terminology from the data analysis domain and consider the purpose and effect of the operations.
Your instructions just need to clearly describe the conversion chain without describing additional operations.
Your instruction should follow the format:  
Instruction: [Your data preparation instruction]
\end{Verbatim}
\end{tcolorbox}

\hypertarget{prompt-step2}{}
\noindent
\begin{tcolorbox}[colback=blue!10,colframe=blue!50!black, breakable, title=Prompt for Instruction Refinement, label=prompt-step2]
\begin{Verbatim}[breaklines=true]
Based on the following data preparation task description, generate a natural language statement expressing the user's intent. 
Concise, Action-Oriented Language: Focus on the core actions and remove unnecessary details. Keep the language clear and direct to highlight the transformation intent.
Clarification of Key Tables and Columns: Maintain essential table names and columns, but express them in a natural, straightforward way.
Simplified Descriptions of Complex Steps: Emphasize the main objectives (sorting, filtering, deduplication) without diving into excessive details, unless they are crucial for the context.
Necessary details need to be preserved such as the suffix of the join, the way the de-duplication operation is performed (first or last), etc.

Here are some examples:
---
Task Description: To transform the input tables into the target table, follow these steps: 
1. Begin by performing an inner join between table_1 and table_2 using the allergy name column from table_1 and the allergy column from table_2. This will combine records from both tables where there is a match on these columns, while including the allergy name and allergy type from table_1 along with the stuid from table_2. 
2. Next, group the resulting dataset by the allergy name (now included in the joined table) and aggregate the data by counting the number of unique stuid entries for each allergy name. This will give you the total number of students associated with each allergy. 
3. After aggregating, sort the grouped data first by the count of stuid in ascending order and then by allergy name in descending order. This will organize the data based on the number of students and the names of the allergens. 
4. From the sorted data, select the top 7 entries based on the highest counts of students. This step ensures that we focus only on the most significant allergens. 
5. Rename the columns in the resulting dataset by changing allergy name to allergen and stuid to student ID to make the column names more intuitive. 
6. Apply a filter to retain only those records where the student ID (which now represents the count of students) is greater than or equal to 3. This will help in identifying the allergens that have a notable number of students associated with them. 
7. Remove any duplicate entries from the filtered dataset to ensure that each allergen-student ID combination is unique. 
8. Finally, perform a sort on the deduplicated data by student ID in ascending order and allergen in descending order to achieve the desired final format. Following these steps will yield a table that lists allergens along with the count of students associated with each, structured as specified in the target table.
User Intent: Start by performing an inner join between table_1 and table_2 on 'allergy name' and 'allergy', with suffixes '_left' and '_right'. Then, group the data by allergy name and count the number of 'stuid' entries for each allergen to determine the number of students associated with each allergy. After grouping, sort the data first by the student count in ascending order and then by allergy name in descending order. Select the top 7 entries. Rename the columns to change allergy name to allergen and stuid to student ID for clarity. Apply a filter to keep only the records where the student ID is 3 or greater. Deduplicate the data, keeping the first occurrence of each duplicate entry to ensure uniqueness. Finally, sort the deduplicated dataset by student ID in ascending order and allergen in descending order to produce the final result.
---
Task Description: First, combine the two input tables, table_1 and table_2, by performing a union operation to consolidate all records, including duplicates. Next, pivot the resulting table to reorganize the data, setting the station names (STN_NAM) as the index, the data provider (DATA_PVDR) as the columns, and using the minimum longitude (LONGITUDE) as the values. After pivoting, rename the column STN_NAM to Station Name. Then, filter the table to keep only the rows where the data provider is "NAV CANADA". Following this, remove any rows that contain missing values in the "NAV CANADA" column. Convert the data type of the "NAV CANADA" column to string. Next, ensure there are no rows where "NAV CANADA" is equal to itself (this condition might be meant for data cleansing or error checking). Finally, deduplicate the entries based on the "NAV CANADA" column while keeping the last occurrence of each duplicate. The result will be your target table with the columns DATA_PVDR and NAV CANADA.
User Intent: Begin by performing a union operation on table_1 and table_2 to consolidate all records, including duplicates. Then, pivot the resulting table with the station names (STN_NAM) as the index, the data provider (DATA_PVDR) as the columns, and use the minimum longitude (LONGITUDE) as the values. Rename the STN_NAM column to "Station Name" for clarity. Next, select the only column "NAV CANADA", and remove any rows with missing values in the "NAV CANADA" column. Convert the "NAV CANADA" column to a string data type and ensure that there are no rows where "NAV CANADA" is equal to itself. Finally, deduplicate the data based on the "NAV CANADA" column, keeping the last occurrence of each duplicate entry.
---
Task Description: First, reshape the data from the wide format to a long format by selecting the columns related to 'PUZZLE B' and 'PUZZLE A', while keeping the specified index columns intact. After transforming the data to a long format, you can apply the explode operation. This operation will split any column containing comma-separated values into individual rows. Next,transforms data from wide format to long format,it keeps the columns in id_vars unchanged and stacks the values from value_vars ("PUZZLE A" and "PUZZLE B") into two new columns: one for the variable names and another for the values.
User Intent: First,reshape the data by collapsing columns that start with "PUZZLE B" or "PUZZLE A" into a long format, while keeping the specified index columns ("Index", "Where are we?") unchanged. The original suffixes from the column names are extracted into a new column called var, using a space as the separator and matching suffixes with a word character pattern (\w+).Then, Explode the "PUZZLE B" column to create separate rows for each puzzle listed, ensuring that each puzzle is split by commas first. Next,transforms data from wide format to long format,the columns specified in id_vars ("Index", "Where are we?") remain unchanged and serve as identifiers for each row. The values in value_vars ("PUZZLE A" and "PUZZLE B") are then stacked into two new columns: one for the variable names and another for the values.
---

Now, based on the following task description, generate a user intent statement:
Transformation Chain: {transform_chain}
Task Description: {task_instruction}

Please output only the intent statement, without explanation or numbering.
\end{Verbatim}
\end{tcolorbox}

\hypertarget{prompt-step3}{}  
\noindent
\begin{tcolorbox}[colback=blue!10,colframe=blue!50!black, title={Prompt for Instruction Verify}, label=prompt-step3]
\begin{Verbatim}[breaklines=True]
Task Background: The user has generated an initial natural language description from a transformation chain, and then used an LLM to generate a user intent statement based on that initial description.
1. **Transformation Chain**: {transform_chain_str}
2. **Initial Natural Language Description**: {instruction}
3. **Generated Intent**: {intent_text}

Task Requirement: Assume you are a data preparation expert. Based on the current intent, can you infer the correct conversion chain, including the details of the parameters?

Output Requirements:
- If the intent allows you to infer a complete and reasonable transformation chain, output:
{{
"is_valid": "true",
"intent": "{intent_text}"
}}
- Otherwise, output:
{{
"is_valid": "false",
"intent": "[Rewritten Intent]"
}}

Please return the result in strict JSON format with no additional explanations.
\end{Verbatim}
\end{tcolorbox}

\subsection{Prompts for Zero-shot LLMs}
\label{zero-shot-prompt}
This section provides the detailed prompt template designed for zero-shot large language models (LLMs). The prompt is carefully constructed to guide the model in generating accurate and semantically faithful instructions without any fine-tuning. It incorporates schema information and explicit instructions to improve model understanding and output quality. This prompt serves as the basis for consistent evaluation across different zero-shot LLMs.

\begin{tcolorbox}[colback=blue!10,colframe=blue!50!black, breakable,title=Prompts for Zero-shot LLMs]
\begin{Verbatim}[breaklines=true]
You are a data expert with extensive knowledge in data preparation pipelines.  
Your task is to select operators based on user intent and use them to transform the source tables.  

Important notes:
- After selecting the operators, ensure they can be correctly executed, especially keeping variable names consistent.
- Note: Except for the `join` and `union` operations, the result table name remains the same as the source table name. For `join` and `union`, the result table name should follow the format `table_x_table_y_join` or `table_x_table_y_union`.

Below are the available operators:
{
  "operators": [
    {
      "name": "join",
      "pandas_equivalent": "merge",
      "parameters": {
        "left_table": "left_table_name",
        "right_table": "right_table_name",
        "result_table": "table_x_table_y_join",
        "left_on": "left_column",
        "right_on": "right_column",
        "how": "",
        "suffixes": ["", ""]
      },
      "description": "Merge two datasets on a common column with specified input/output table names"
    },
    {
      "name": "union",
      "pandas_equivalent": "concat",
      "parameters": {
        "source_tables": ["table_1", "table_2"],
        "axis": 0,
        "result_table": "table_x_table_y_union",
        "ignore_index": true,
        "how": ["all", "distinct"]
      },
      "description": "Vertically concatenate multiple tables (similar to SQL UNION)"
    },
    {
      "name": "groupby",
      "pandas_equivalent": "groupby",
      "parameters": {
        "source_table": "source_table_name",
        "group_by": ["group_column_1", "group_column_2"],
        "aggregations": {
          "value_column_1": "aggregation_function",
          "value_column_2": "aggregation_function"
        },
        "result_table": "source_table_name"
      },
      "description": "Group data by specified columns and apply aggregation (similar to SQL GROUP BY)"
    },
    {
      "name": "pivot",
      "pandas_equivalent": "pivot_table",
      "parameters": {
        "source_table": "source_table_name",
        "index": ["index_column_1", "index_column_2"],
        "columns": ["column_to_expand"],
        "values": ["value_column"],
        "aggfunc": "aggregation_function",
        "result_table": "source_table_name"
      },
      "description": "Convert long-format data into wide-format (similar to Excel Pivot Table)"
    },
    {
      "name": "unpivot",
      "pandas_equivalent": "melt",
      "parameters": {
        "source_table": "source_table_name",
        "id_vars": ["fixed_column_1", "fixed_column_2"],
        "value_vars": ["column_to_unpivot_1", "column_to_unpivot_2"],
        "var_name": "variable",
        "value_name": "value",
        "result_table": "source_table_name"
      },
      "description": "Convert wide-format data into long-format (similar to SQL UNPIVOT)"
    },
    {
      "name": "explode",
      "pandas_equivalent": "pd.explode",
      "parameters": {
        "source_table": "source_table_name",
        "result_table": "source_table_name",
        "column": "list_column",
        "split_comma": True or false
      },
      "description": "Expand column values into separate rows (separate them by commas first if necessary)"
    },
    {
      "name": "filter",
      "pandas_equivalent": "query",
      "parameters": {
        "source_table": "source_table_name",
        "condition": "`column_name` operation value",
        "result_table": "source_table_name"
      },
      "description": "Filter rows based on conditions (similar to SQL WHERE)"
    },
    {
      "name": "sort",
      "pandas_equivalent": "sort_values",
      "parameters": {
        "source_table": "source_table_name",
        "by": ["column_1", "column_2"],
        "ascending": [true, false],
        "result_table": "source_table_name"
      },
      "description": "Sort data by specified columns (similar to SQL ORDER BY)"
    },
    {
      "name": "wide_to_long",
      "pandas_equivalent": "pd.wide_to_long",
      "parameters": {
        "source_table": "source_table_name",
        "subnames": ["subname"],
        "i": ["id_column"],
        "j": "var",
        "sep": "",
        "suffix": "",
        "result_table": "source_table_name"
      },
      "description": "Convert wide-format data to long-format"
    },
    {
      "name": "transpose",
      "pandas_equivalent": "transpose",
      "parameters": {
        "source_table": "source_table_name"
      },
      "description": "Transpose rows and columns of a table; no additional parameters needed"
    },
    {
      "name": "rename",
      "pandas_equivalent": "rename",
      "parameters": {
        "source_table": "source_table_name",
        "rename_map": "Dictionary mapping old column names to new names"
      },
      "description": "Rename columns based on the provided mapping"
    },
    {
      "name": "dropna",
      "pandas_equivalent": "dropna",
      "parameters": {
        "source_table": "source_table_name",
        "subset": ["List or single column name to check for missing values"],
        "how": "Deletion strategy: either 'any' or 'all'"
      },
      "description": "Remove rows with missing values in specified columns"
    },
    {
      "name": "deduplicate",
      "pandas_equivalent": "drop_duplicates",
      "parameters": {
        "source_table": "source_table_name",
        "subset": ["List or single column name to determine duplicates"],
        "keep": ["first", "last"]
      },
      "description": "Remove duplicate rows, keeping either the first or last occurrence in each group"
    },
    {
      "name": "topk",
      "pandas_equivalent": "head(k)",
      "parameters": {
        "source_table": "source_table_name",
        "k": "Number of top rows to retain"
      },
      "description": "Select the top k rows after sorting by index or specific criteria"
    },
    {
      "name": "select",
      "pandas_equivalent": "loc / bracket selection",
      "parameters": {
        "source_table": "source_table_name",
        "columns": "List of column names to keep"
      },
      "description": "Select specified columns from the table"
    },
    {
      "name": "cast",
      "pandas_equivalent": "astype",
      "parameters": {
        "source_table": "source_table_name",
        "column": "Column name to change data type",
        "dtype": "Target data type (e.g., 'int', 'float', 'str')"
      },
      "description": "Convert the data type of the specified column"
    }
  ]
}

Please output the transformation steps from the input tables to the target table using the above operations.  
The output should follow the JSON format below:
```json
[
  {
    "name": "",
    "parameters": {
    }
  }
]
User Intent:  
{USER_INTENT}  
Input Tables (First 10 Rows):  
{SOURCETABLE}  
Please reason step-by-step based on the user intent, and then provide the result.  
The final output should be in JSON format.
"""
\end{Verbatim}
\end{tcolorbox}

\newpage
\subsection{Prompts for Text-to-Pandas}
\label{text-to-pandas-prompt}

To ensure accurate generation of pandas code from natural language instructions, we designed structured prompt template tailored to various data preparation operations. The template guide the model in understanding user intent, interpreting table schemas, and producing syntactically and semantically correct code.

\begin{tcolorbox}[colback=blue!10,colframe=blue!50!black, breakable,title=Prompt for Text-to-Pandas]
\begin{Verbatim}[breaklines=true]
I need you to convert natural language into Pandas code.

dataset schema:
{dataset_schema}

question: {task['question']}

Code Return Guidelines:
1. If you need to return a DataFrame as the result, assign it to a variable named 'result'
2. If you modify an existing DataFrame, keep its original variable name
3. If you create a new DataFrame (other than the final result), use a clear variable name (e.g., df_temp)
4. Each input table is already loaded as a DataFrame. The variable name of each DataFrame is the same as the input table name, such as table_1, table_2, etc.

For join operations:
1. Use the appropriate merge/join method based on the requirement
2. Make sure to specify the correct 'on' or 'left_on'/'right_on' parameters
3. Use the appropriate join type (inner, left, right, outer) as required
4. After joining, assign the result to the 'result' variable.

Please generate Pandas code that solves this problem. Only return the code, no explanation. 
Ensure the code is executable and follows the return guidelines above.

\end{Verbatim}
\end{tcolorbox}

\subsection{Prompts for Text-to-SQL}
\label{text-to-sql-prompt}

To support accurate and consistent SQL generation, we design structured prompt templates as following that pair natural language instructions with table schemas.

\begin{tcolorbox}[colback=blue!10,colframe=blue!50!black, breakable,title=Prompt for Text-to-SQL]
\begin{Verbatim}[breaklines=true]
I need you to convert natural language questions into SQL queries.
        
The database schema is as follows:
{schema_prompt}

question: {sample["question"]}

Please generate an SQL query that can answer this question. Only return the SQL query, no explanation is needed. Ensure your SQL query is executable in SQLite.
\end{Verbatim}
\end{tcolorbox}

\section{Visualization Platform}
\label{app:platform}
We develop an interactive visualization platform to help human-experts inspect and debug data, featuring a dashboard and task-level panels for description, code, and table comparison.

\begin{figure}[htbp]
    \centering
    \includegraphics[width=0.5\textwidth]{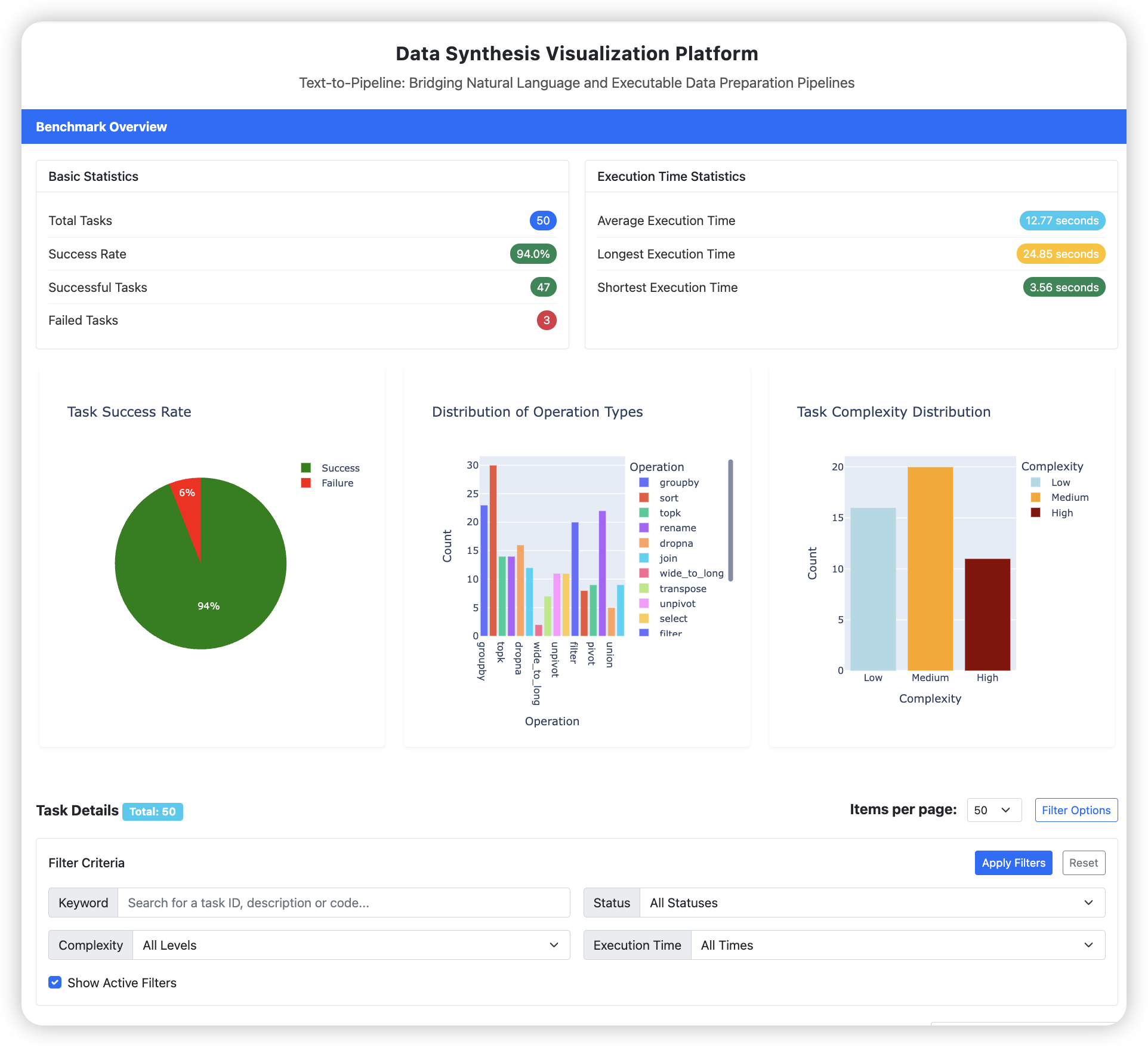}
    \caption{Benchmark Overview. The dashboard summarizes total task count, success rate, execution time statistics, operator distributions, and task complexity.}
    \label{fig:platform-overview}
\end{figure}

\vpara{Dashboard Overview}
This overview dash board~(Fig.~\ref{fig:platform-overview}) displays metadata for each task, including status, execution time, predicted complexity, and involved operations (e.g., \texttt{groupby}, \texttt{sort}, \texttt{topk}).

\begin{figure}[h]
    \centering
    \includegraphics[width=0.5\textwidth]{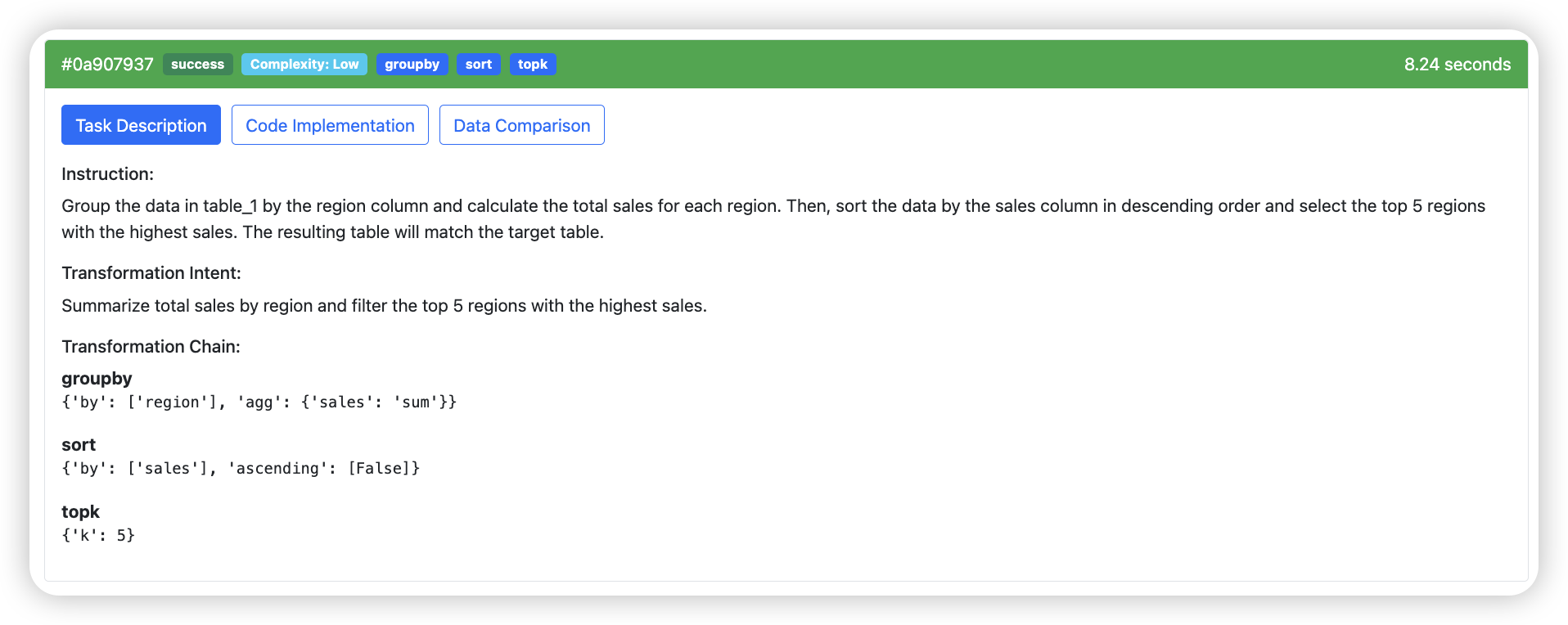}
    \caption{Task Description Panel. It displays the original natural language instruction, the rewritten instruction, and a structured transformation chain in DSL.}
    \label{fig:platform_task_desp}
\end{figure}

\vpara{Task Description Panel} 
This panel~(Fig.~\ref{fig:platform_task_desp}) shows the original natural language instruction, its interpreted transformation intent, and the corresponding symbolic transformation chain.

\begin{figure}[h]
    \centering
    \includegraphics[width=0.5\textwidth]{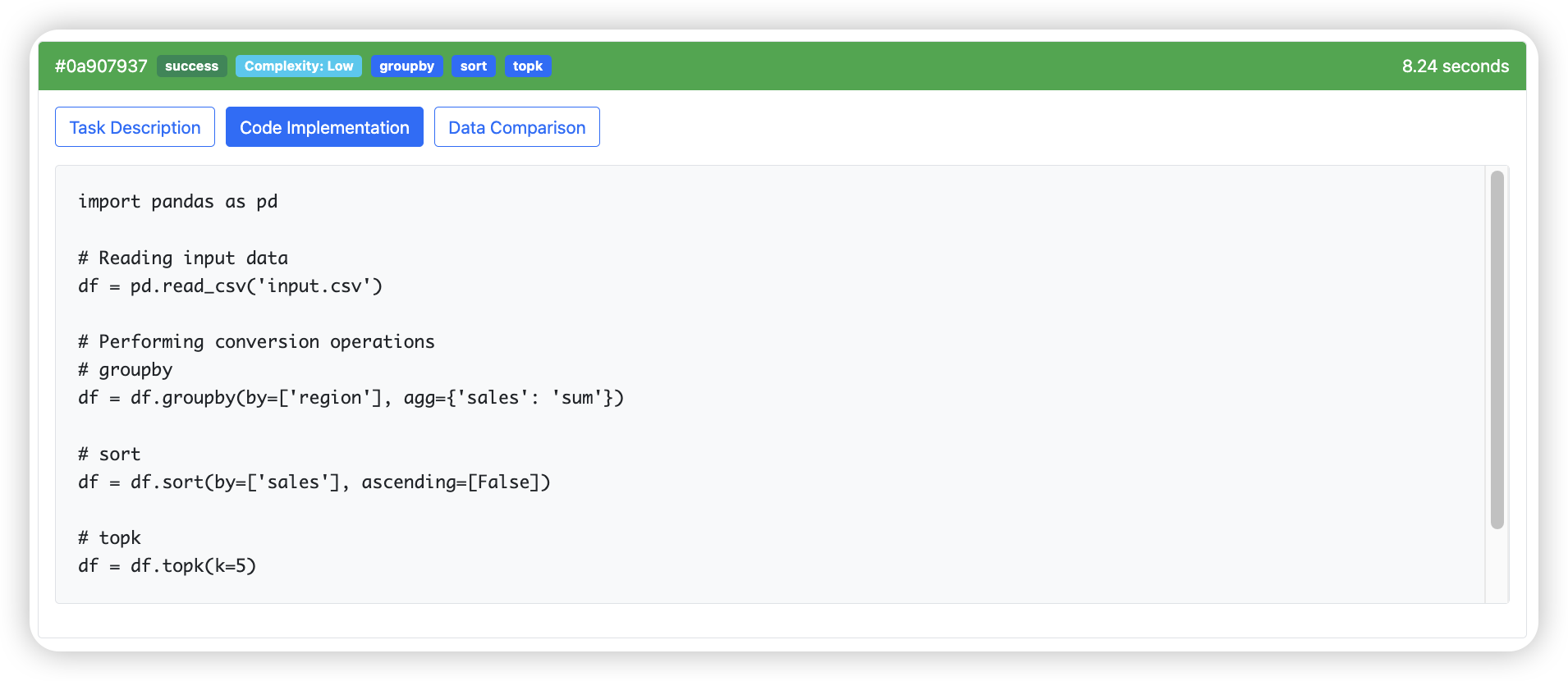}
    \caption{Code Implementation Panel. It presents the synthesized Python (Pandas) code that executes the DSL logic.}    
    \label{fig:platform_code_impl}
\end{figure}

\vpara{Code Implementation Panel}
This panel~(Fig.~\ref{fig:platform_code_impl}) presents the compiled Python (Pandas) implementation generated from the symbolic program.

\begin{figure}[h]
    \centering
    \includegraphics[width=0.5\textwidth]{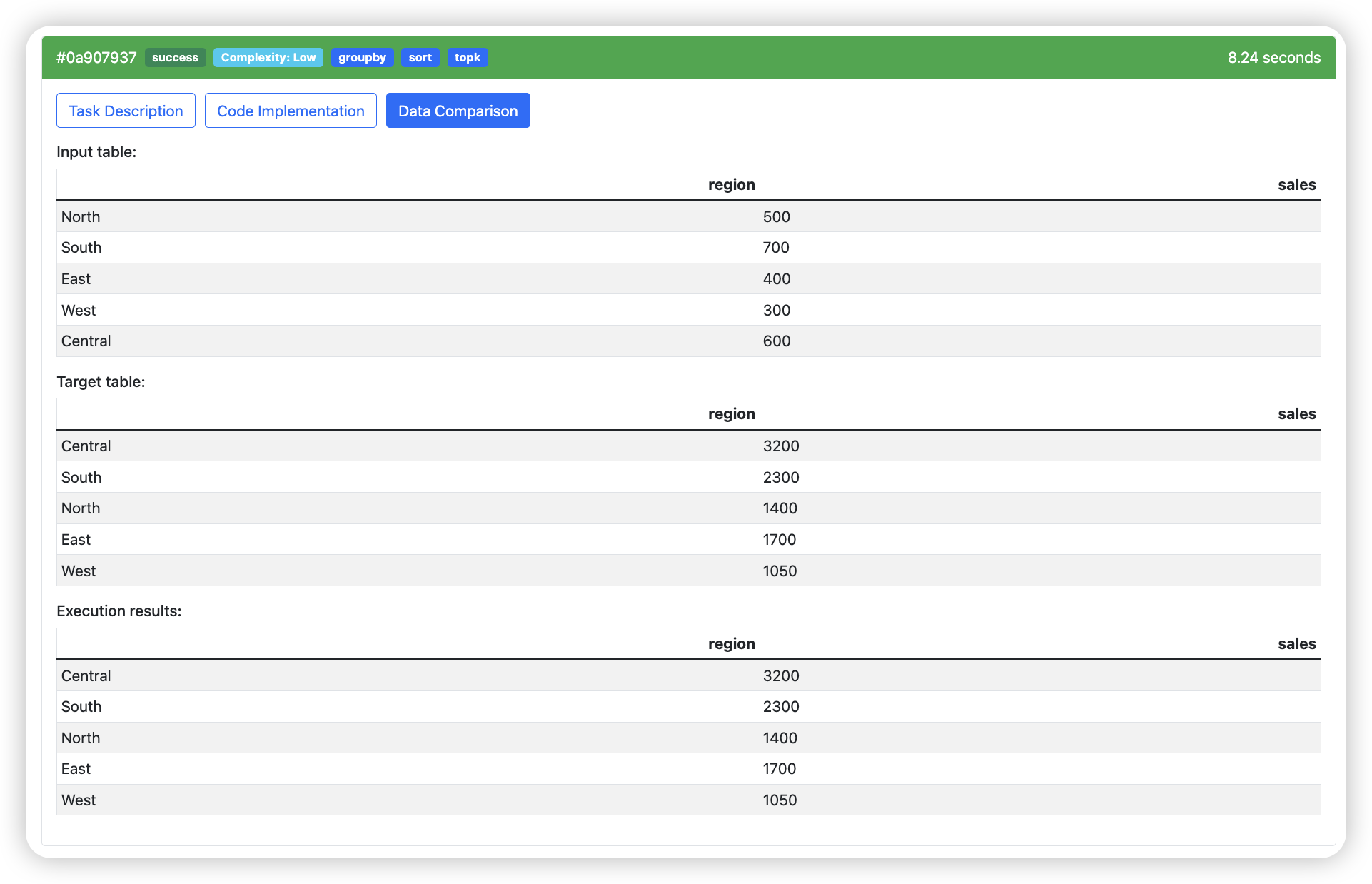}
    \caption{Data Comparison Panel. It visualizes the input table, target output, and actual execution result side-by-side for verification.}    
    \label{fig:platform_table_compasion}
\end{figure}

\vpara{Table Comparison Panel}
This panel~(Fig.~\ref{fig:platform_table_compasion}) provides side-by-side visualization of the input table, ground-truth output, and actual model execution result.
\par\vspace{0.5em}

\end{document}